\documentclass[aps,prb,preprint,groupedaddress]{revtex4-1}

\usepackage{mathrsfs}
\usepackage{amssymb}
\usepackage{amsmath}
\usepackage{enumitem}
\usepackage{bm}
\usepackage[unicode=true, bookmarks=true,bookmarksnumbered=true,bookmarksopen=false, breaklinks=false,pdfborder={0 0 0},backref=false,colorlinks=true,citecolor=magenta]{hyperref}
\usepackage{graphicx}
\usepackage{showlabels}
\usepackage{multirow}

\def\<{\langle}
\def\>{\rangle}
\def\O{{\cal O}}
\newcommand{\ordine}[1]{{\sf O}[#1]}

\begin{document}


\title{The low-temperature phase in the two-dimensional long-range diluted XY model}



\author{Fabiana Cescatti$^{1}$}
\author{Miguel Ib\'a\~nez-Berganza$^{4}$}
\email{miguel.berganza@roma1.infn.it}
\author{Alessandro Vezzani$^{3}$}
\author{Raffaella Burioni$^{1,2}$}
\affiliation{$^1$Dipartimento di Scienze Matematiche, Fisiche e Informatiche, Universit\`a di Parma, via G.P. Usberti, 7/A - 43124, Parma, Italy}
\affiliation{$^2$INFN, Gruppo Collegato di Parma, via G.P. Usberti, 7/A - 43124, Parma, Italy}
\affiliation{$^3$IMEM-CNR, Parco Area delle Scienze, 37/A-43124 Parma, Italy}
\affiliation{$^4$Dipartimento di Fisica, Universit\`a di Roma, ``La Sapienza''. Piazzale Aldo Moro, 5, 00185 Roma, Italy.}


\date{\today}

\begin{abstract}
The critical behaviour of statistical models with long-range interactions exhibits distinct regimes as a function of $\rho$, the power of the interaction strength decay. For $\rho$ large enough, $\rho>\rho_{\rm sr}$, the critical behaviour is observed to coincide with that of the short-range model. However, there are controversial aspects regarding this picture, one of which is the value of the short-range threshold $\rho_{\rm sr}$ in the case of the long-range XY model in two dimensions. We study the 2d XY model on the {\it diluted} graph, a sparse graph obtained from the  2d lattice by rewiring links with probability decaying with the Euclidean distance of the lattice as $|r|^{-\rho}$, which is expected to feature the same critical behavior of the long range model. Through Monte Carlo sampling and finite-size analysis of the spontaneous magnetisation and of the Binder cumulant,  we present numerical evidence that $\rho_{\rm sr}=4$. According to such a result, one expects the model to belong to the Berezinskii-Kosterlitz-Thouless (BKT) universality class for $\rho\ge 4$, and to present a $2^{nd}$-order transition for $\rho<4$.

\end{abstract}


\maketitle

\tableofcontents

\section{Introduction \label{sec:intro}}

The two-dimensional XY model is one of the most relevant models in statistical mechanics since it describes the Berezinskii-Kosterlitz-Thouless (BKT) phenomenology, the paradigm of topological phases \cite{kosterlitz2017} which cannot be described by conventional symmetry breaking and long-range order. The BKT phenomenology \cite{jos2013}  is exhibited by a wide variety of physical systems in two dimensions: film superfluids\cite{bishop1978} and superconductors,\cite{epstein1981,benfatto2007} superconducting arrays (see references in \cite{fazio2001}), cold atomic systems,\cite{hadzibabic2006,murthy2015} and quantum many body systems in $1+1$ dimensions,\cite{giamarchi2004} among others.  

Moreover, the XY model exhibits an important theoretical interest {\it per se}. It is used as a playground for the theories of critical phenomena,\cite{pelissetto2005,defenu2017-2} of disordered systems,\cite{binder1986,coolen2005,marruzzo2016,lupo2018} of the chirality transition,\cite{alba2010,obuchi2013} of fractional dimensions,\cite{codello2015} of long-range interactions,\cite{defenu2015,defenu2017} of dynamical synchronisation\cite{gupta2014} and of phase transitions in complex networks.\cite{burioni1999,dorogovtsev2008,berganza2013}

A question of particular theoretical interest regarding the XY model is the possible generalisation of the BKT phenomenology on the regular two-dimensional (2d) lattice with long-range interactions decaying with the Euclidean distance of the lattice as $|r|^{-\rho}$.\cite{dauxois2009} The question is framed in a more general topic regarding the behaviour of statistical models in $d$-dimensions with long range interactions, that  has been the subject of a vast literature. The critical behaviour of these models is characterised by the value of $\rho$: they may either belong to the mean field universality class (for $\rho\le\rho_{\rm mf}$), or to a non-mean field universality class with $\rho$-dependent critical exponents (for $\rho_{\rm mf}\le \rho\le\rho_{\rm sr}$), or to the short-range, $d$-dimensional reference lattice universality class (for $\rho\ge \rho_{\rm sr}$), where $\rho_{\rm mf}$, $\rho_{\rm sr}$ are $d$- and model-dependent numbers. It is known, indeed, that increasing the long-range character of the interaction (by decreasing $\rho$) in the long-range $O(n)$ model on a $d$-dimensional lattice, leads to a critical behaviour that is equivalent to that of a short-range model in an {\it equivalent fractional dimension}, $D_{\rm e}(\rho)\ge d$.\cite{defenu2015,codello2015} In particular, the self-consistent relation for $D_{\rm e}$ is: $D_{\rm e}(\rho)=2[d-\eta_{\rm sr}(D_{\rm e}(\rho))]/(\rho-d)$, where $\eta_{\rm sr}(D)$ is the anomalous dimension of the short-range model in $D$ fractional dimensions. This relation is known as the {\it short-long range} equivalency. While in the limit $n\to\infty$ it is exact, for finite $n$ there exist several controversial aspects of the long--short range mapping (see references in \cite{angelini2014}), and the validity of the expression for $D_{\rm e}(\rho)$ near the short-range threshold $\rho\searrow\rho_{\rm sr}$ is still openly debated.

One of the most debated question in this context is the value of $\rho_{\rm sr}$. Fisher et. al.\cite{fisher1972} first put forward the presence of the three regimes: for $d<\rho\le 3d/2$, the critical behaviour coincides with that of the mean-field model; for $3d/2<\rho\le \rho_{\rm sr}$, the system exhibits non-mean field, peculiar critical exponents depending on $\rho$; finally, for $\rho > \rho_{\rm sr}$, the critical behavior is expected to coincide with that of the short-range model in $d$ dimensions (recovered for $\rho\to\infty$). They proposed that $\rho_{\rm sr}=2+d$ and that, for $3d/2<\rho\le \rho_{\rm sr}$, the anomalous dimension is $\eta(\rho)=2+d-\rho$. In this context, at $\rho=\rho_{\rm sr}=d+2$, $\eta(\rho)$ exhibits a discontinuity from $0$ to $\eta_{\rm sr}=\eta(\infty)$, i.e. the value of anomalous dimension in the short range model. 

However, Sak\cite{sak1973} introduced a different picture in which $\eta(\rho)=2+d-\rho$ is continuous and the long-short range transition arises when $\rho_{\rm sr}=2+d-\eta_{\rm sr}$ (the value for which $\eta(\rho)=\eta_{\rm sr}$). This picture was confirmed by Monte Carlo (MC) simulations for the Ising model,\cite{luijten2002} and contrasted by subsequent numerical work\cite{picco2012,blanchard2013} which proposed a non-mean field correction to $\eta(\rho)$ recovering $\rho_{\rm sr}=2+d$, as in Fisher's work. Recently, further numerical work\cite{angelini2014} has shown  that logarithmic corrections undermine the numerical estimation of the critical exponents in ref. \cite{picco2012,blanchard2013} and they propose Sak's theory as the simplest explanation of the observed results. Moreover, Sak's result  $\rho_{\rm sr}=2+d-\eta_{\rm sr}$ has been recovered by using the functional renormalization group beyond the local potential approximation,\cite{defenu2015} the equivalence of the free energy finite size scaling,\cite{leuzzi2008,leuzzi2009,banos2012,angelini2014} and perturbation theory.\cite{honkonen1989,honkonen1990}

The value of $\rho_{\rm sr}$ in the case of the  two-dimensional XY model ($n=2$, $d=2$) is particularly controversial. A blind application of the functional renormalisation group provides, for the anomalous dimension of the two-dimensional XY model, $\eta_{\rm sr}(n,d)=0$ and, hence, $\rho_{\rm sr}=4$.\cite{defenu2015,codello2013} However, at the BKT transition $\eta^{\rm (BKT)}_{\rm sr}=1/4$ and, consequently, the application of Sak's formula would give $\rho_{\rm sr}=3.75$, which is consistent with the numerical results of ref.\cite{berganza2013} We notice that, for the two-dimensional XY model, $\rho_{\rm sr}$ separates two completely different critial behaviours: for $\rho>\rho_{\rm sr}$, a topological phase transition takes place, while for $\rho<\rho_{\rm sr}$, it is expected a standard transition to a magnetic phase.

If the correct value was $\rho_{\rm sr}=3.75$, the system would present a peculiar phase for $\rho\in (3.75:4)$. As a matter of fact, it is known that  the existence of a magnetised low-temperature phase for continuous symmetry models on a graph is determined by its spectral dimension, $d_s$: the generalized Mermin-Wagner theorem and its inverse proved in refs.\cite{cassi1992,burioni1999} indeed state that the XY model presents spontaneous magnetization if and only if $d_s > 2$. The value of  $d_s$ has been calculated in \cite{burioni1997} for a two-dimensional lattice with long range interactions showing that $d_s>2$ as soon as $\rho<4$. As a consequence, the long-range two-dimensional XY model is expected to exhibit spontaneous magnetisation at under-critical temperatures for $\rho<4$, in agreement with further exact results.\cite{picco1983,kunz1976,defenu2015,codello2015} Therefore, $\rho_{\rm sr}=3.75$  would imply that for $3.75<\rho<4$ the system would be both magnetised and it would feature some characteristics of the BKT class, making the interval $\rho\in (3.75:4)$ particularly interesting.

An important point is that much about the critical behaviour of the long range XY model has been studied when it is defined on the {\it long-range diluted graph}, a sparse graph obtained from a regular lattice with links rewired with probability decaying according to the same power $\rho$ of the interaction decay. Interestingly, the diluted graph  interpolates between the {\it reference} (regular) {\it lattice} in $d$ dimensions, for $\rho\to \infty$, and a random graph, for $\rho \to 0$.\cite{leuzzi2008}  Moreover, the numerical study in ref.\cite{berganza2013} suggests that the long-range diluted graph has the same spectral dimension as its fully connected long range counterpart.

It is, hence, expected that $(\rho,d)$-long range diluted $O(n)$ models in $d$ dimensions  present the same universality class as their fully connected equivalents with long-range interactions, decaying with the same power $\rho$ of the distance in a $d$-dimensional lattice.\cite{leuzzi2008,leuzzi2009,katzgraber2009,banos2012,angelini2014} This mapping is {\it believed} to hold, it is known to be at least approximately valid and it has been studied in \cite{berganza2013} in the case of the 2d XY model. Therefore, the long-range diluted graph provides a valid testing ground to investigate the nature of the XY low temperature phase and the value of $\rho_{\rm sr}$.

In this article we study the long-range diluted  XY model and we provide rather unambiguous numerical evidence of the fact that in the interval $3.75 < \rho <4$, the system: (1) presents spontaneous magnetisation, and (2) is not scale invariant for under-critical temperatures. These results confirm the generalised Mermin-Wagner direct and inverse theorems\cite{cassi1992,burioni1999}  and suggests, in agreement with \cite{defenu2015}, that the long range XY model belongs to the BKT universality class as far as $\rho>\rho_{\rm sr}=4$.

In the following section we will define the model. In section \ref{sec:methods} we will present the numerical protocol that we have used, and the objectives of our article. Section \ref{sec:results} is to present the results, while our conclusions are drawn in section \ref{sec:conclusions}.

\section{The long-range diluted XY model \label{sec:context}}

\begin{figure}
\includegraphics[width=.325\columnwidth]{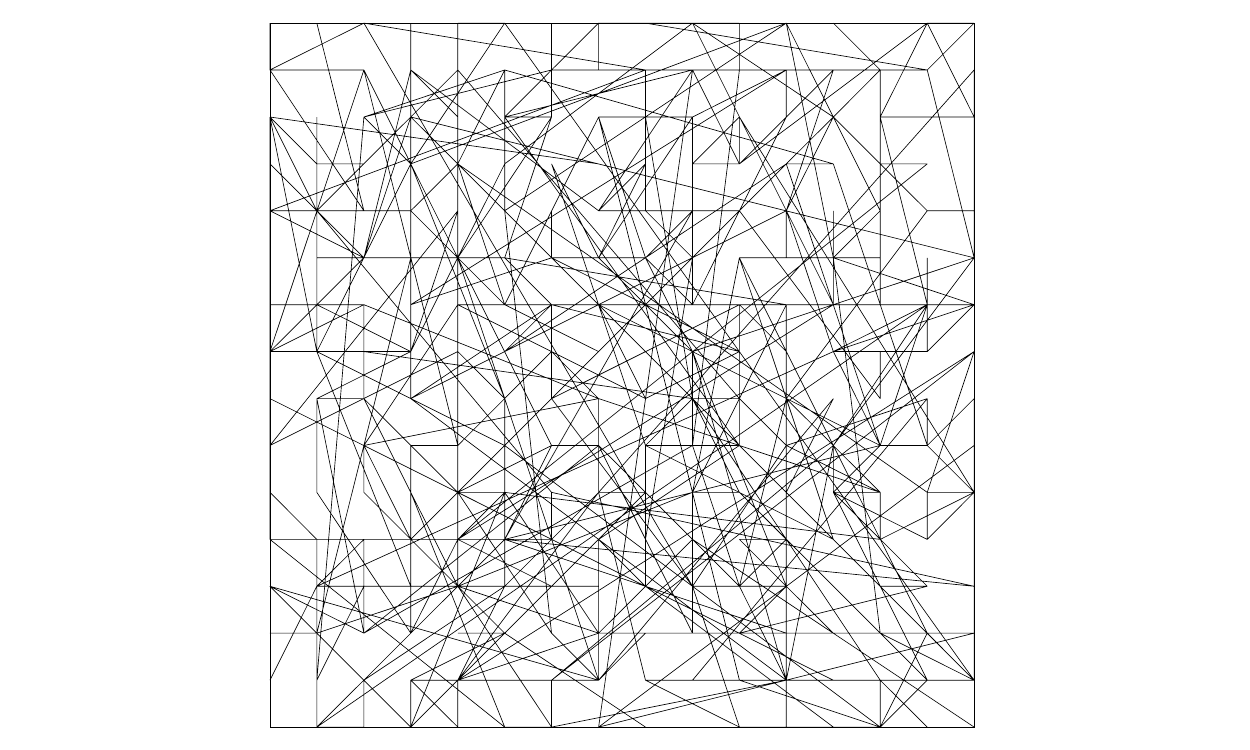}
\includegraphics[width=.325\columnwidth]{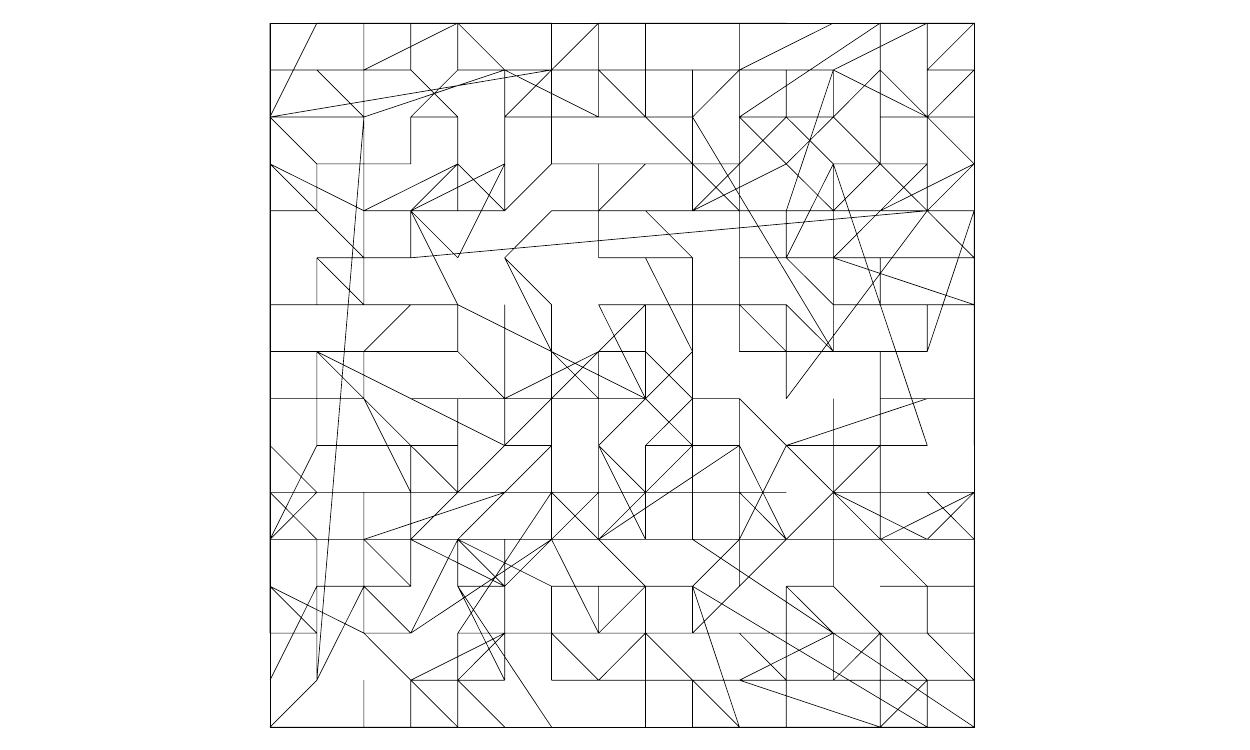}
\includegraphics[width=.325\columnwidth]{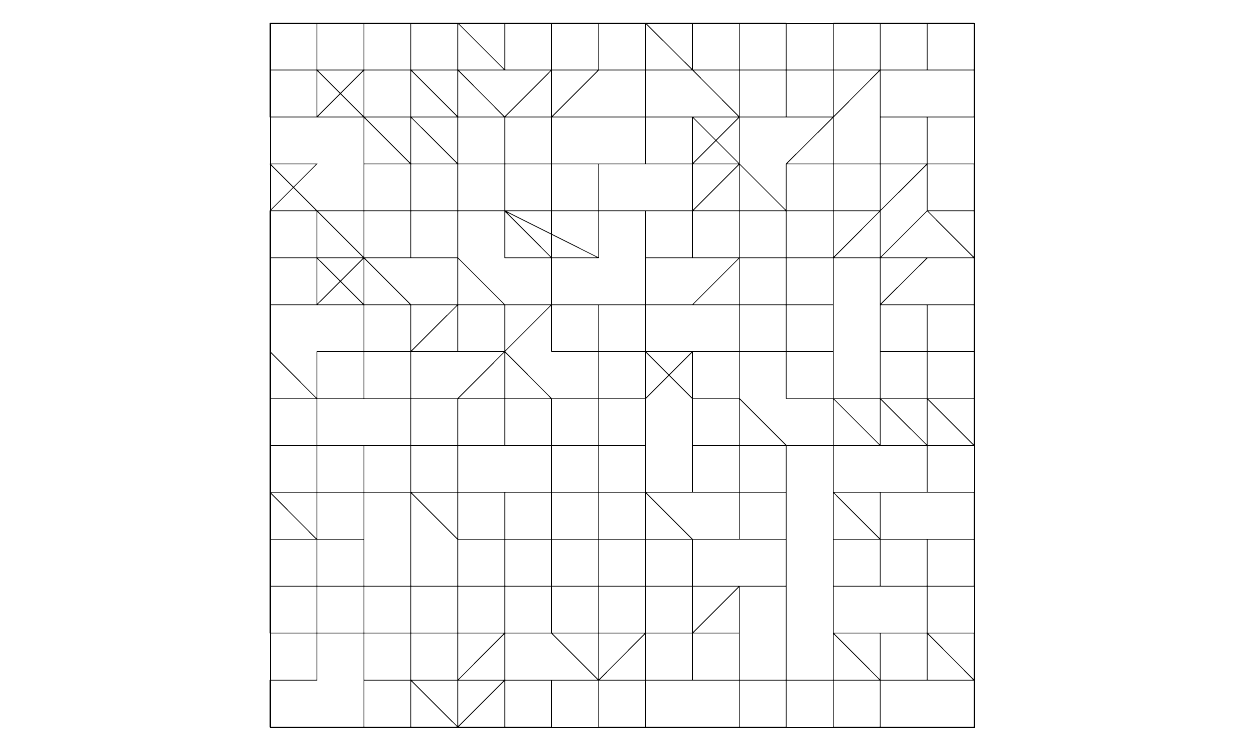}
	\caption{ A graphical illustration of three realisations of the long-range diluted graph. The reference lattice is a $d=2$, $N=16^2$, square lattice with free boundary conditions, with $\rho=2$, $4$, and $8$ (from left to right). The total number of links is constant, $N_l=2(N-N^{1/2})$.  \label{fig:dilutegraphs}}
\end{figure}

The long-range diluted XY  model is defined by the Hamiltonian:

\begin{equation}
	{\cal H}[\{{\bf s}_i\}] = -\sum_{i<j} A_{ij} {\bf s}_i \cdot {\bf s}_j
\end{equation}
where ${\bf s}_j$, $j=1,\ldots,N$, is the degree of freedom of the $j$-th site, is a two-dimensional vector with unit norm, and where $A$ is the adjacency matrix, $A_{i,j}=0,1$, of an undirected network that belongs to the ensemble of long-range dilute graphs. They are such that the probability of a link between two sites $i,j$, $A_{ij}=1$ is proportional to $|{\bf r}_{ij}|^{-\rho}$, being $|{\bf r}_{ij}|$ its distance in a {\it reference} hypercubic $d$-dimensional lattice, constrained such that the number of links in the graph is a fixed, extensive number, $N_l=\sum_{i<j=1}^N A_{ij}$, being $N_l=2N$ for periodic boundary conditions (for details on the numerical construction of this graph see \cite{berganza2013} and appendix \ref{sec:graphgeneration}). When $\rho\to\infty$, only nearest neighbors of the original $d$-dimensional lattice survive since their probability is infinitely larger than that of longer range links, while, in the limit $\rho\to 0$, the probability of a given link does not depend on ${\bf r}_{ij}$, and the ensemble of graphs coincides with the \"Erdos-R\'enyi ensemble with $N$ nodes and $N_l$ links. In fig. \ref{fig:dilutegraphs} we show three realizations of long-range dilute graphs for  $d=2$, $N=16^2$, and three values of $\rho$.

As we already mentioned in the introduction, it is believed that the critical behaviour of $(d,\rho)$-diluted models coincides with that of their long-range equivalents at the same value of $d$ and $\rho$. Indeed, in ref.\cite{berganza2013}, for the XY model in $d=2$ the three regimes (mean field, $\rho$-dependent non-mean field, short range) have been found, with a strong numerical evidence for $\rho_{\rm mf}=3d/2$ and a moderate numerical evidence for the more subtle estimation $\rho_{\rm sr}\simeq 3.75$. 
However, the question of the presence of spontaneous magnetisation has not been directly addressed in ref. \cite{berganza2013}.

As explained in the introduction, the XY model is known to present spontaneous magnetisation only when defined in graphs of spectral dimensions $d_{\rm s} > 2$.\cite{cassi1992,burioni1999} Both for the long range and the diluted graph the spectral dimension is known to be:\cite{burioni1997,berganza2013}

\begin{eqnarray}
	d_{\rm s}=\infty,	\qquad	&& \rho\le d \\
	d_{\rm s}=\frac{2d}{\rho-d},	\qquad &&	d<\rho\le 2+d  \\
	d_{\rm s}=d, \qquad	 &&		d\ge 2+d 
	\label{eq:spectraldimension}
	.
\end{eqnarray}
According to this result, we expect to find spontaneous magnetisation in the $d=2$ long-range dilute XY model as far as $\rho<4$, and no magnetized phase when $\rho>4$.
This fact may induce to propose the following excluding possibilities. (A) $\rho_{\rm sr}=3.75$ and in the interval $\rho\in (3.75:4)$ the model presents an anomalous behavior, with some traits in common with the BKT phase but, at the same time, with spontaneous magnetization. (B) $\rho_{\rm sr}=3.75$ and in the interval $\rho\in (3.75:4)$ the model present standard BKT transition. In this case the inverse Mermin-Wagner extension\cite{burioni1999} would not hold for some reason in long range systems. (C) $\rho_{\rm sr}=4$: the BKT phase disappears at the same value of $\rho$ at which a magnetic state arise. 

This provides a motivation for the study of the critical behavior for $\rho\in (3.75:4)$. We will show that our numerical results are compatible with the hypothesis (C) since the system at low temperature presents spontaneous magnetization and no scale invariance, confirming  the generalized inverse Mermin-Wagner theorem.\cite{burioni1999}

\section{Methods\label{sec:methods}}
We will numerically address, by means of Monte Carlo  sampling and finite size scaling, whether the long-range diluted 2d XY model presents spontaneous magnetization and whether it presents scale invariance for $\rho$ in the interval $(3.75:4)$. For this purpose we will focus on the case $\rho=3.875$ in the middle point of the interval under study. For this value of  $\rho=3.875$, according to Eq. (\ref{eq:spectraldimension}),  $d_{\rm s} \simeq 2.1333$. As a reference comparison, we have also considered $\rho=4.5$, for which the system is expected to belong to the BKT universality class and to present no spontaneous magnetisation at low temperatures.

\subsection{ Fluctuations and error estimation.}

Two different sources of fluctuations are presents in the average of a generic observable $\O$: first, the thermal fluctuations, $[\<\O^2\>-\<\O\>^2]$; second, the fluctuations due to the different  realizations of the graph topology, $[\<\O\>^2]-[\<\O\>]^2$, where $\<\cdot\>$ represents the ensemble average for a given realization of the graph, and $[\cdot]$ represents the average over the ensemble of long-range diluted graphs with power $\rho$. As error bars in our numerical calculations, we have considered the latter, since it is, by definition, larger (see appendix \ref{sec:errors}).  The calculation of both quantities can be used to evaluate the relative influence of inter-graph fluctuations with respect to thermal fluctuations within a given graph.

Algorithmically, the ensemble average $\<\cdot\>$ is estimated by the average $\<\cdot\>_{\rm MC}$ over successive configurations of the MC Markov chain with a given initial condition, sequence of random numbers and graph realisation, while $[\cdot]$ is estimated by $[\cdot]_{\rm MC}$, the average over different realizations of the Markov chain and the random graph (see further details in appendix \ref{sec:errors}).  The MC dynamics provides correlated sequences of observable estimations. The correct error estimations can be computed through the Jack-Knife method, which accounts for such a correlation (see \cite{amit2005} and Appendix \ref{sec:errors}). From now onwards, we will refer to $[\cdot]_{\rm MC}$ and $\<\cdot\>_{\rm MC}$ simply as $[\cdot]$ and $\<\cdot\>$.

\subsection{The observables}

{\bf Spontaneous magnetisation.\label{sec:absm}}  The spontaneous magnetisation will be evaluated as the average $[\<|{\bf m}|\>] (N,\beta)$ of the observable:

\begin{equation}
	{\bf m}(\{{\bf s}_j\}) =  \frac{1}{N} \sum_{j=1}^N {\bf s}_j
\label{eq:magnetization}
\end{equation}

As we explain in appendix \ref{sec:errors}, all the observables depending explicitly on the components of the magnetization ($m_{x}$, $m_{y}$) present a high correlation time, much longer than the simulation time, while this is not the case for the observables depending only on its modulus $|{\bf m}|$ (see appendix \ref{sec:errors}). 
The absolute value of the magnetisation $[\<|{\bf m}|\>]$ is, hence, a suitable quantity to numerically address whether a system presents spontaneous magnetisation for under-critical temperatures. A system without spontaneous magnetisation in the thermodynamic limit, $\lim_{N\to \infty}[\<|{\bf m}|\>] (N,\beta)=0$, would exhibit a residual spontaneous magnetisation at finite sizes, $[\<|{\bf m}|\>] (N,\beta)>0$ that is, however, expected to decrease as a homogeneous power law for large $N$: $[\<|{\bf m}|\>] (N,\beta) \sim N^{-a(\beta)}$. Vice-versa, a system presenting spontaneous magnetisation would present a residual magnetisation in the limit of large $N$: $\lim_{N\to \infty} [\<|{\bf m}|\>] (N,\beta)>0$. As a numerical strategy to address the presence/absence of spontaneous magnetisation, we will check the absence/presence of an homogeneous power law dependence on $N$ of $[\<|{\bf m}|\>] (N,\beta)$. 

{\bf Scale invariance.} The question of the presence of scale invariance is addressed by means of the analysis of the Binder cumulant:

\begin{equation}
	B(N,\beta)=2-\left[\frac{\<{\bf m}^4\>}{\<{\bf m}^2\>^2}\right] \label{eq:bindercumulant}
\end{equation}
In the infinitely large $N$-limit, the Binder cumulant converges to $1$ and to $0$ in the low and hight-temperature phases, respectively. Systems presenting a second-order phase transition exhibit a crossing of the $B(N,\beta)$ curves in the $\beta$-axis for various values of $N$. The crossing of the $B(N,\beta)$ and $B(N/2,\beta)$ curves provides an estimation of the finite-size critical temperature $\beta^*(N)$ that, in the large-$N$ limit, converges with power-law corrections in $N$ to a single critical point, $\beta^*$.\cite{amit2005} Oppositely, a system in the BKT universality class is expected to present scale invariance for under-critical temperatures, hence the Binder cumulant is expected to converge to a $\beta$-dependent value for large $N$, $B(\infty,\beta)<1$  {\it for all values of} $\beta>\beta^*(N)$ (see, for example, refs. \cite{ballesteros2000,berganza2013}); one expects a superposition of different $B(N,\beta)$ curves for different, sufficiently large  $N$'s and for all $\beta>\beta^*$, rather than a crossing at $\beta^*$. 

{\bf Other observables.} We have also considered other thermodynamic quantities, mainly: the susceptibility, $\chi(N,\beta)=N[\<{\bf m}^2\>-\<{\bf m}\>^2]$; the susceptibility of the modulus of the magnetization,  $\chi_{|{\bf m}|}(N,\beta)=N[\<{\bf m}^2\>-\<|{\bf m}|\>^2]$; the argument of the magnetisation vector $[\<\arctan(m_y/m_x)\>]$; and the single components of the magnetisation, $[\<m_x\>]$. In appendix \ref{sec:errors} we present some result regarding the correlation time of these observables. However, our conclusions about the spontaneous magnetisation and the scale invariance are drawn from the study of $[\<|{\bf m}|\>] (N,\beta)$ and $B(N,\beta)$. We will  present them in the following section.

\subsection{Numerical protocol}

{\bf Description of the simulations.} For our analysis we have employed Monte Carlo  sampling in the canonical ensemble at inverse temperature $\beta$, using the Metropolis algorithm.\footnote{We have deliberatly avoided the use of other thermalisation algorithms, as the Parallel Tempering algorithm, to be able to perform an analysis of our algorithm correlation time for each observable, treating all the sizes $L$ on the same ground (that could have been done only tuning the parallel tempering inter-temperature spacing in such a way that the exchange rate is equal for all the sizes $L$.}
 We have simulated various values of $\beta$ in the interval $[0.7:1.1)$, in particular: $\beta=0.7+k(1.1-0.7)/32$, $k=0,\ldots,31$, and various values of the linear size $L=N^{1/2}$ of the reference lattice (a square lattice with periodic boundary conditions): $L=2^\ell$ with $\ell=6,\ldots,10$, along with shorter simulations at two further values of $L=192$, $384$. 

For each temperature, we have simulated an $L$-dependent number $N_{r}(L)$ of realisations of a diluted graph. For each one, we have performed an $L$-dependent number of MC sweeps devoted to observable sampling, $n_{\rm MCS}(L)$, preceded by a further $1/8$-th of $n_{\rm MCS}(L)$ MC sweeps devoted to the equilibration from the initial condition. The components of the global magnetisation (equation (\ref{eq:magnetization})) of each configuration have been saved in memory every $t_{\rm s}=16$ sweeps (so that we have a total of $n_{{\rm meas}}(L)=n_{\rm MCS}(L)  N_{r}(L)/t_{\rm s}$ measures for each observable and inverse temperature). In table \ref{tab:TL} we report $n_{\rm MCS}$ and $N_r$ for the various simulations.  

\begin{table}
\begin{small}
\caption{\label{tab:TL}}
\begin{ruledtabular}
\begin{tabular}{ccccccc}
        &  \multicolumn{3}{c}{$\rho=3.875$}& \multicolumn{3}{c}{$\rho=4.5$}\\
        \cline{2-4}\cline{5-7}
	$L$ & $N_r(L)$  & $n_{MCS}(L)$ & $n_{{\rm meas}}$ & $N_r(L)$  & $n_{MCS}(L)$ & $n_{{\rm meas}}$\\
	\hline
	$64$ & $20$ & $1835008$ & $2293760$ & $20$ & $1835008$ & $2293760$ \\
	$128$ & $20$ & $1835008$ & $2293760$ & $20$ & $1835008$ & $2293760$ \\
	$256$ & $20$ & $1835008$ &  $2293760$ & $20$ & $1835008$ & $2293760$ \\
	$512$ & $15$ & $1835008$ & $1720320$ &  $20$ & $1835008$ & $2293760$ \\
	$1024$ & $15$ & $1835008$ & $1720320$ & $8$ &  $1835008$ & $917504$ \\
	$192$ & $100$ & $114688$ & $716800$ & $100$ & $114688$ & $716800$\\
	$384$ & $100$ & $114688$ &  $716800$ & $100$ & $114688$ & $716800$
\end{tabular}
\end{ruledtabular}
\end{small}
\end{table}


{\bf Equilibration checks.} Due to the presence of logarithmic corrections near the incriminated threshold $\rho\simeq 4$,\cite{angelini2014} the correlation times of the relevant observables becomes rather large (see appendix \ref{sec:errors}), and the MC thermalisation and calculation of observables is by far  more subtle than that of the square lattice XY model. For this reason, he have paid particular attention to the analysis of equilibration times and to the error estimation of the observables. As equilibration tests, for each of the relevant observables, we have considered the averages as equilibrated as far as the corresponding correlation time, estimated through the Jack-Knife method (see \cite{amit2005} and appendix \ref{sec:errors}) is much lower than the simulation time, and that the averages and standard deviations of the observables  performed over {\it data blocks} of exponentially increasing length, become independent on the block length (see appendix \ref{sec:errors}).

{\bf Software.} To cope with the computational complexity of the problem, we have employed an efficient  Graphics Processing Units software for MC sampling of spin models in arbitrary sparse topologies, developed by the authors of \cite{berganza2013} in 2012, whose performance, reaching below the nano-second per spin flip, was reviewed in ref. \cite{berganza2013}. The software exhibits several improvements with respect to the 2012 version, specially an improved algorithm for the efficient generation of the long-range diluted graph for large values of $\rho$ and $N$, that turns necessary for the simulation of many realizations of the $L=2^{9}$ and $L=2^{10}$ graphs, and that is described in appendix \ref{sec:graphgeneration}.

\section{Results \label{sec:results}}

For both the analysis of $[\<|{\bf m}|\>]$ and of the Binder cumulant $B$, we have considered a $\rho$-dependent set ${\cal B}_\rho$ of $8$ values of $\beta$, which are reported in the legend of fig. \ref{fig:absm}. ${\cal B}_\rho$ is defined by the subset of the simulated inverse temperatures that lie in the interval $[\beta_\rho^*:(0.925)^{-1}\beta_\rho^*]$, i.,e., in an under-critical temperature range whose length is $0.075\, T_\rho^*$. The critical inverse temperature $\beta^*$ for each value of $\rho$ has been roughly estimated by inspection of the Binder cumulant crossing (or overlapping), see sec. \ref{sec:scaleinvariance} and fig. \ref{fig:bindervsbeta}. 

\subsection{Spontaneous magnetisation}

\begin{figure}
\includegraphics[width=.49\columnwidth]{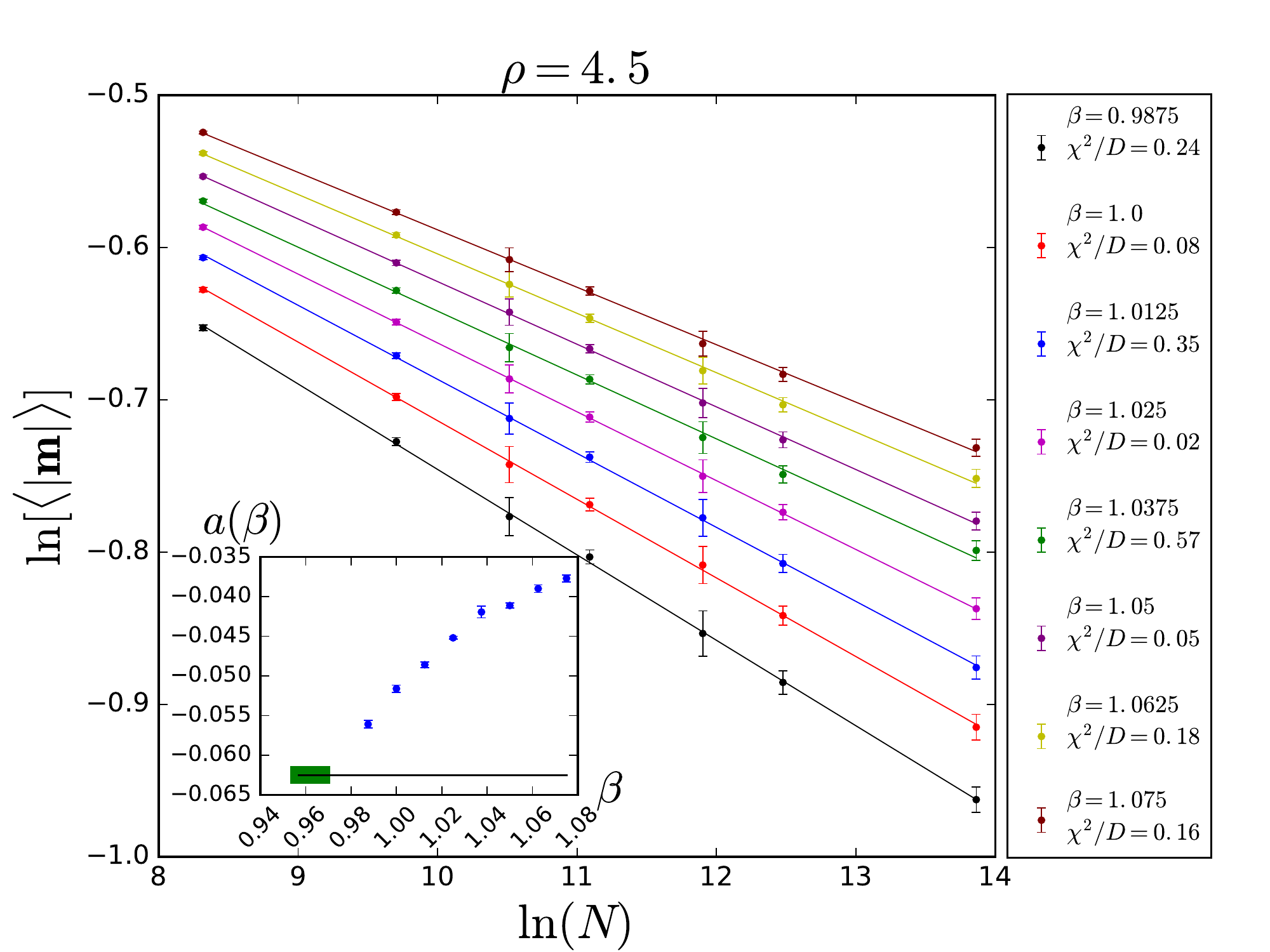}
\includegraphics[width=.49\columnwidth]{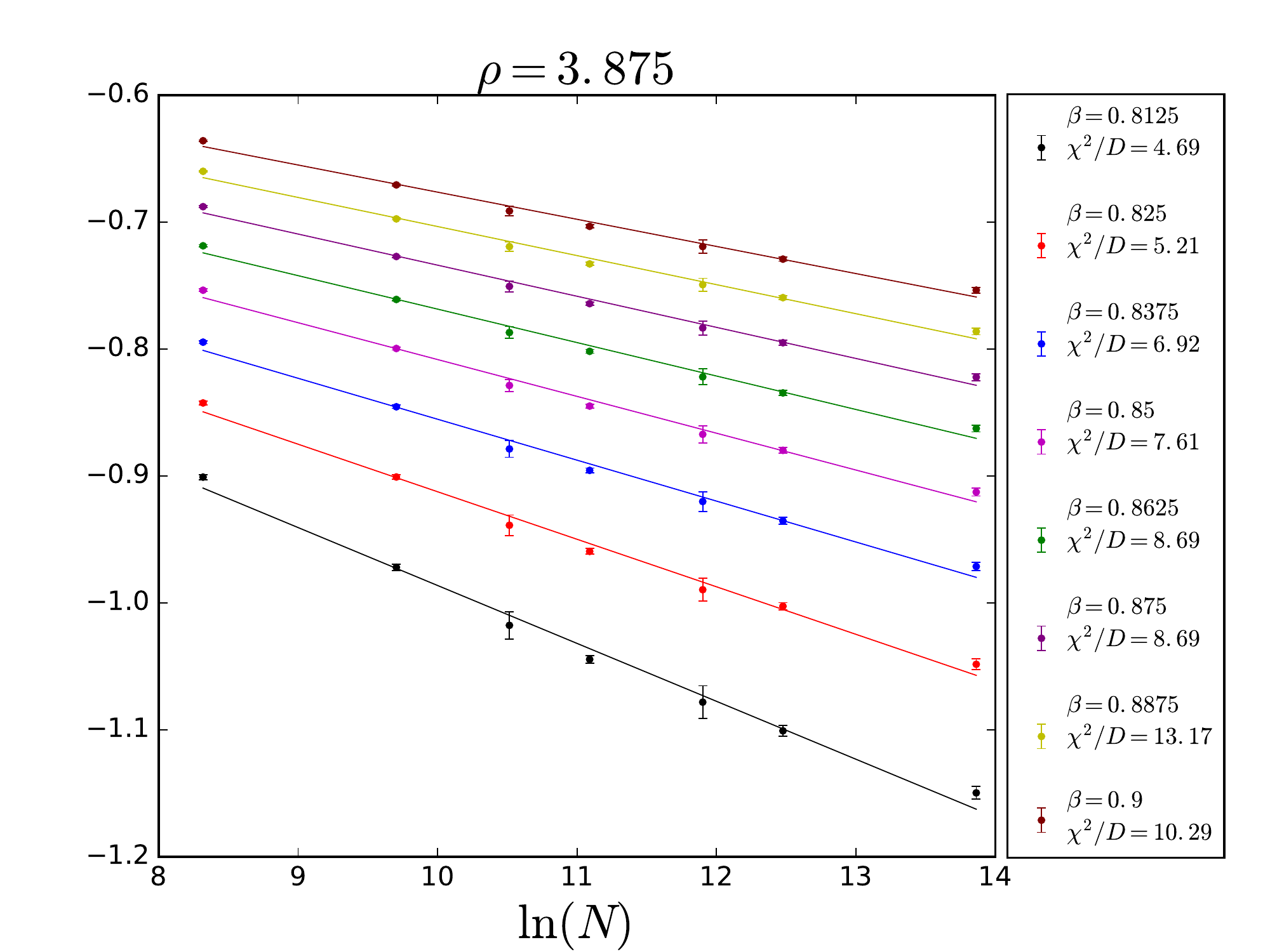}
	\caption{Left: $\ln [\<{|\bf m|}\>](N,\beta)$ versus $\ln N$ for $\rho=4.5$.  Different curves correspond to different values of $\beta>\beta^*_\rho$ in ${\cal B}_{4.5}$ (from bottom to top $\beta$ increases). The lines are linear regressions whose summed squared residuals per degree of freedom are shown in the figure caption for all $\beta$'s. Inset: the estimated exponent $a(\beta)$ vs $\beta$. The horizontal line indicates its predicted value at $\beta^*_{4.5}$, $1/16$, while the thick green segment over the horizontal line indicates the estimated interval for $\beta^*_{4.5}$. Right: idem for $\rho=3.875$. \label{fig:absm}}
\end{figure}

We present the results concerning the quantity $[\<|{\bf m}|\>]$ in fig. \ref{fig:absm}, in which we plot $\ln [\<{|\bf m|}\>](N,\beta)$ versus $\ln N$ for each of the two models, $\rho=4.5$, $\rho=3.875$. In fig. \ref{fig:absm}-Left, we report first the reference value $\rho=4.5$, which we expect to exhibit no spontaneous magnetisation. We find, indeed, a clear power law of the form $\ln [\<|{\bf m}|\>](N,\beta) \simeq -a(\beta) \ln N$ (the sum of squares residuals per degree of freedom of the linear regression is lower than one for all the temperatures $\beta> \beta^*$). The value of the exponent $2a(\beta)$ in the $\rho=4.5$ case is reported in the inset of fig. \ref{fig:absm}-left. Its extrapolation to the estimation interval for $\beta^*_{4.5}$ (estimated as explained in sec. \ref{sec:scaleinvariance}) is compatible with its theoretical expectation value, $a=\eta_{\rm BKT}/4=1/4$. Indeed, in the BKT universality class, $\<{\bf m}^2\>$ is known to depend on $N$ at fixed under-critical temperatures as $\<{\bf m}^2\>(N,\beta)\sim N^{-\eta(\beta)/2}$, with $\eta(\beta)\to\eta_{\rm BKT}$ as $\beta\searrow\beta^*$.\cite{berganza2013} Hence, supposing that the $N$-dependence at fixed $\beta$ of $\<|{\bf m}|\>$ coincides with that of $\<{\bf m}^2\>^{1/2}$, we expect that $a(\beta)$ converges to $1/16$ as far as $\beta\searrow\beta^*$ for $\rho=4.5$ in the short-range regime. 

\begin{figure}[h]
\includegraphics[width=.49\columnwidth]{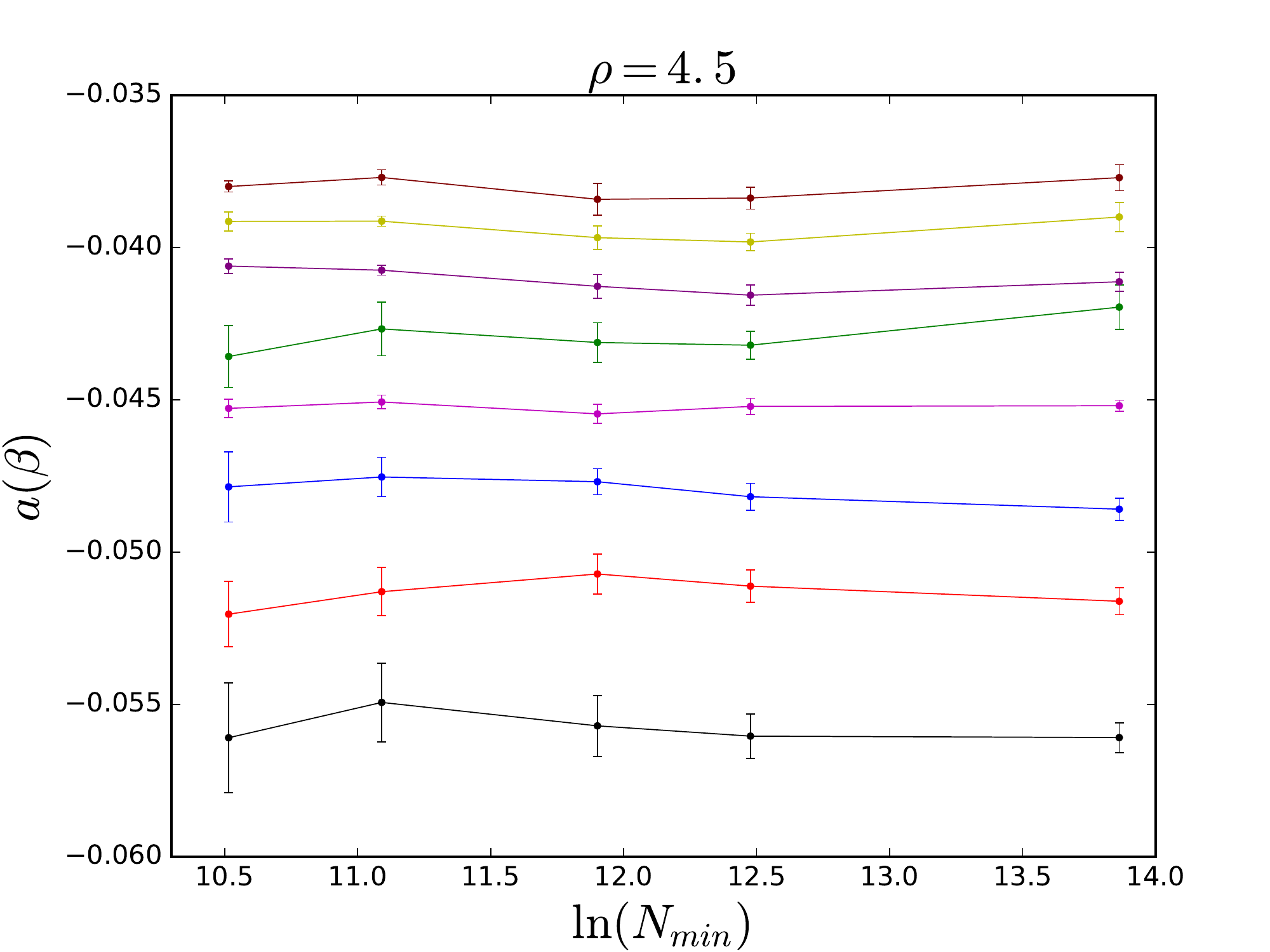}
\includegraphics[width=.49\columnwidth]{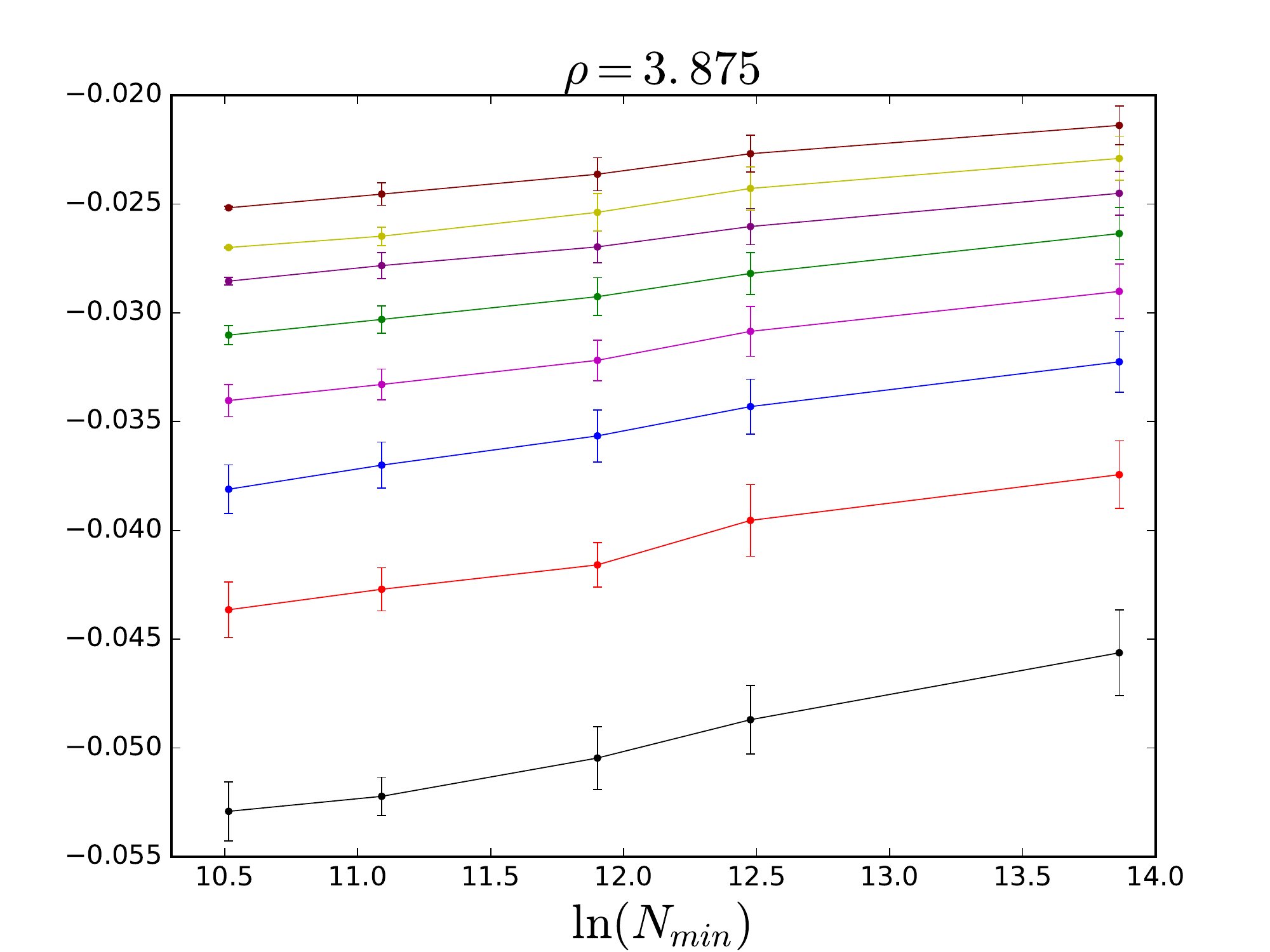}
	\caption{ Left: estimation of $a(\beta)$ from the fits of $\ln [\<{|\bf m|}\>](N,\beta)$ versus $\ln N$ in fig. \ref{fig:absm}, versus the minimum value of the interval used for the fit, $N_{min}$, for $\rho=4.5$. Different curves correspond to different $\beta$'s in ${\cal B}_{4.5}$ (from bottom to top $\beta$ increases). Right: idem, for $\rho=3.875$. \label{fig:absmfit}}
\end{figure}

The case $\rho=3.875$, instead, cannot be fit into a linear power law (the sum of squared residuals per degree of freedom of the linear regression is larger than one for all the temperatures in ${\cal B}_{3.875}$, see the inset of fig. \ref{fig:absm}-right). The dependence of $\ln [\<|{\bf m}|\>](N,\beta)$  on $\ln N$ is {\it concave} for $\rho=3.875$, it decreases {\it slower} than a power law, strongly suggesting the presence of a nonzero homogeneous term, $[\<|{\bf m}|\>](\infty,\beta)>0$. The concavity of $\ln [\<|{\bf m}|\>](N,\beta)$ vs $\ln N$ is more clearly exposed in fig. \ref{fig:absmfit}, where we report the estimation for the exponent $a(\beta)$, through linear regression $\ln [\<|{\bf m}|\>](N,\beta)\sim -a(\beta) \ln N$ of the data in fig. \ref{fig:absm} for different values of  $N_{min}$, the  minimum value of the interval $[N_{min}:1024^2]$ used in the regression (in abscissa). The case $\rho=4.5$ presents a slope independent of $N_{min}$ within its errors (obtained accounting for the standard deviation of the input data in fig. \ref{fig:absm} through orthogonal distance regression\cite{brown1990}). Oppositely, the $\rho=3.875$ case presents a slope which decreases in absolute value with $N_{min}$, indicating that the $\ln [\<|{\bf m}|\>](N,\beta)$ is concave with respect to $\ln N$.

Furthermore, in Appendix \ref{sec:binderNscaling}, we show further numerical evidence of the fact that, in the $\rho=3.875$ case, the average modulus of the magnetisation is compatible with the law: $\<|{\bf m}|\>(N,\beta) = \mu_{\infty}(\beta) + c_{\rho,\beta} N^{-x}$, with $x\simeq 1/8$. 

We conclude that the data regarding the absolute value of the magnetisation strongly suggests that the $\rho=3.875$ dilute XY model exhibits spontaneous magnetisation below the critical temperature, while for $\rho=4.5$ this is not the case, as predicted by the generalized Mermin-Wagner theorem.

\subsection{Scale invariance \label{sec:scaleinvariance}}

\begin{figure}
\includegraphics[width=.49\columnwidth]{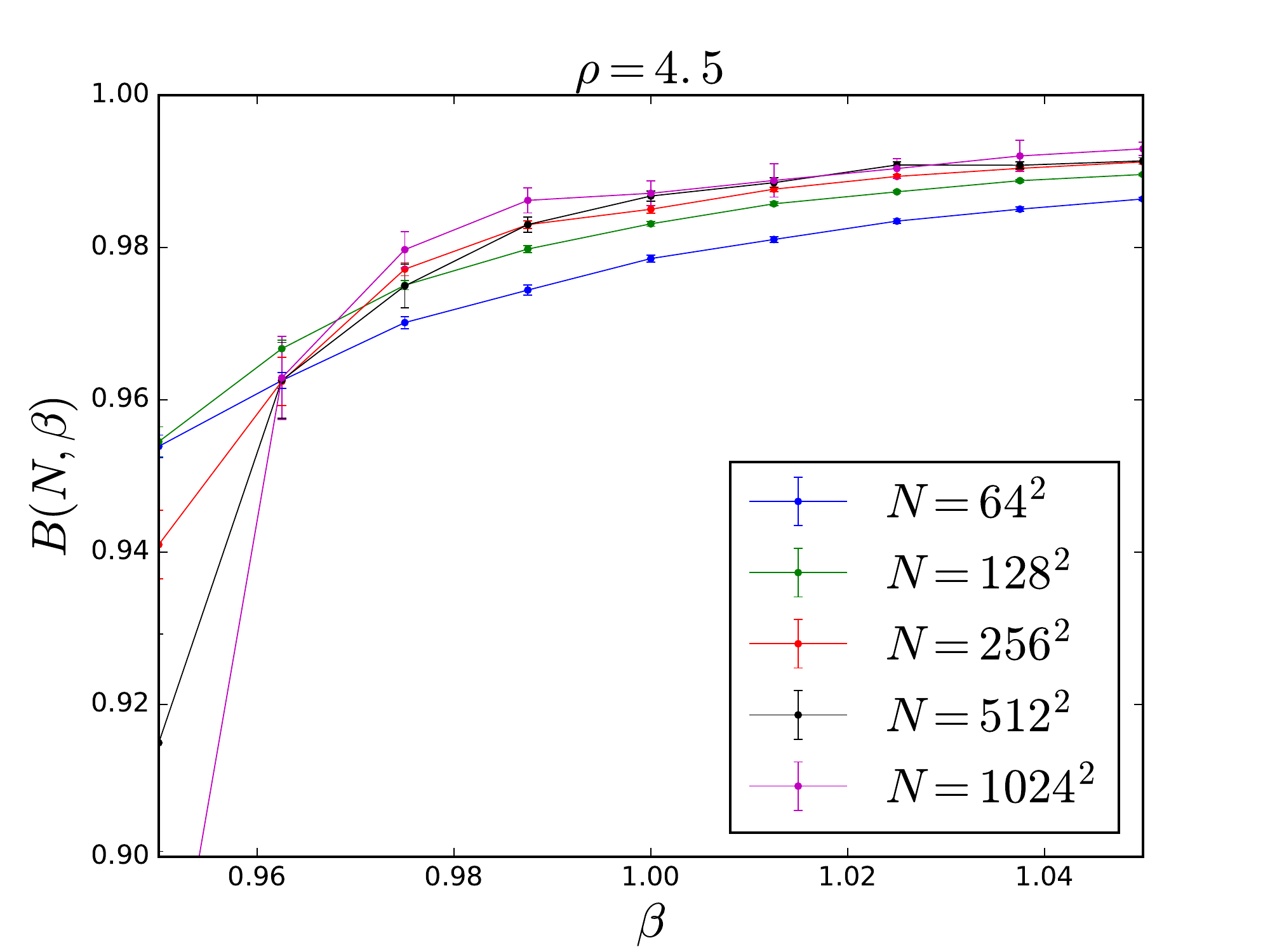}
\includegraphics[width=.49\columnwidth]{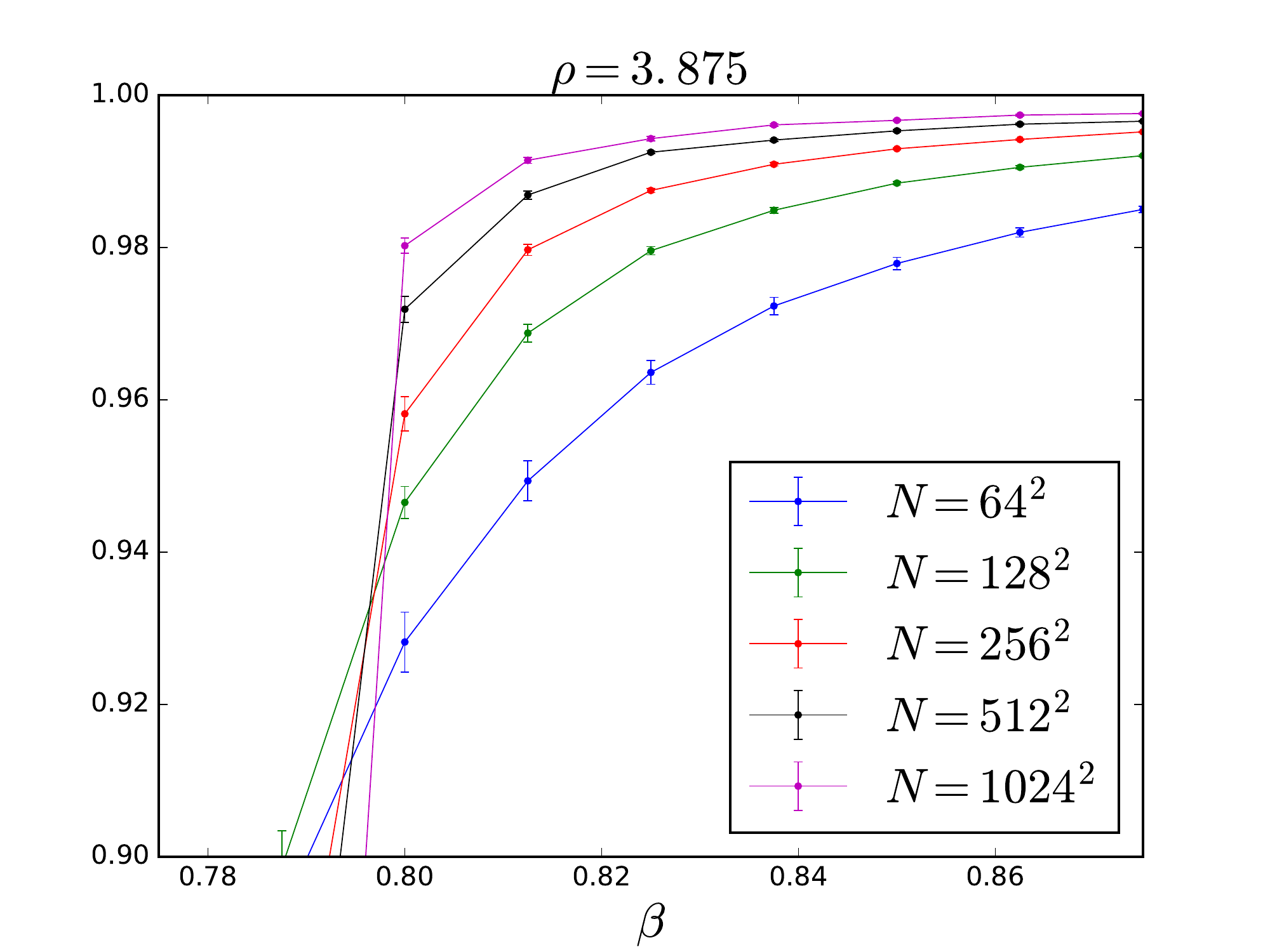}
	\caption{Left: The Binder cumulant $B(N,\beta)$ vs $\beta$ for $\rho=4.5$. Different curves correspond to different values of $N$ (from bottom to top $N$ increases). Lines are a guide to the eye. Right: idem, for $\rho=3.875$. \label{fig:bindervsbeta}}
\end{figure}

\begin{figure}
\includegraphics[width=.49\columnwidth]{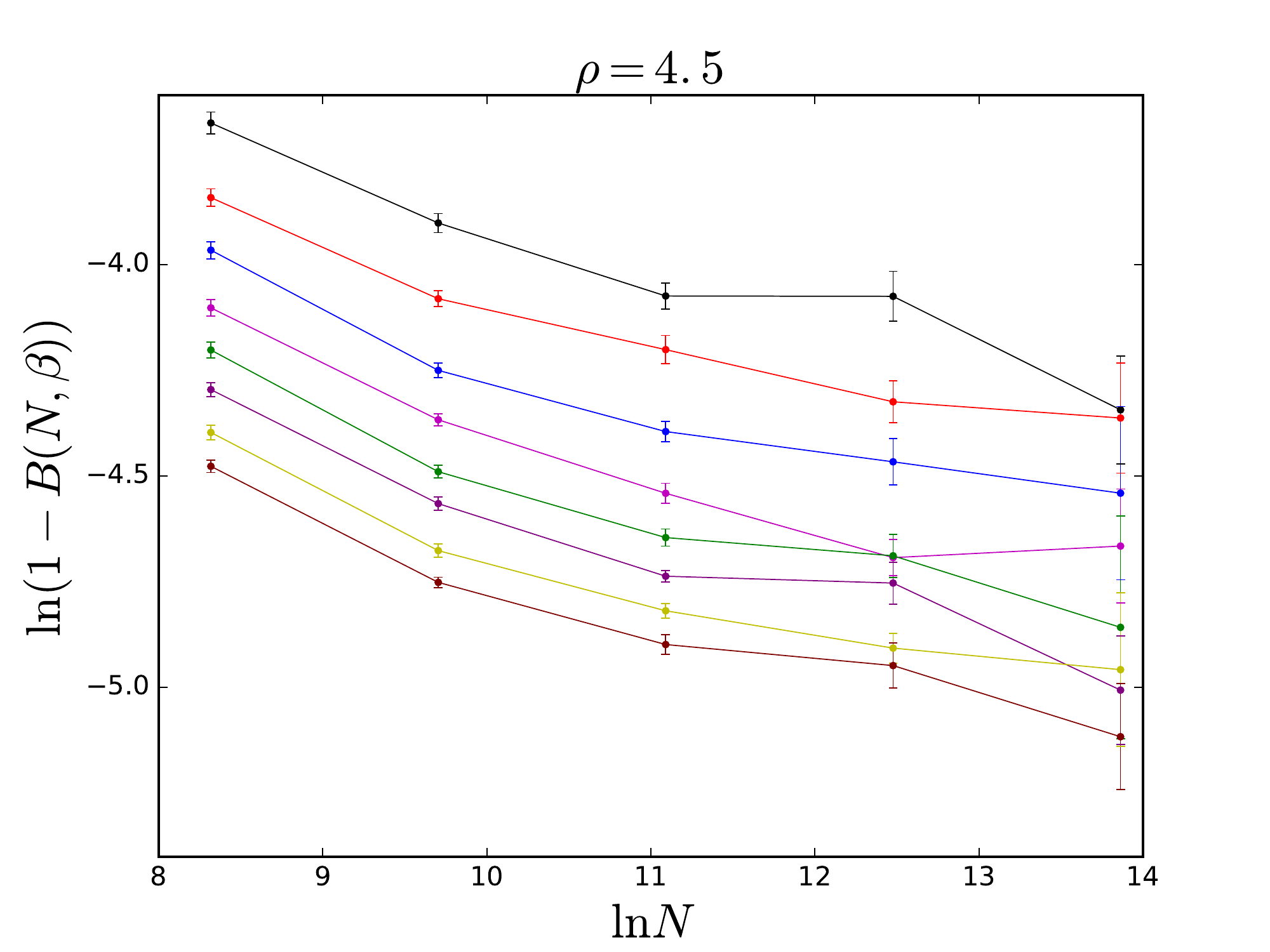}
\includegraphics[width=.49\columnwidth]{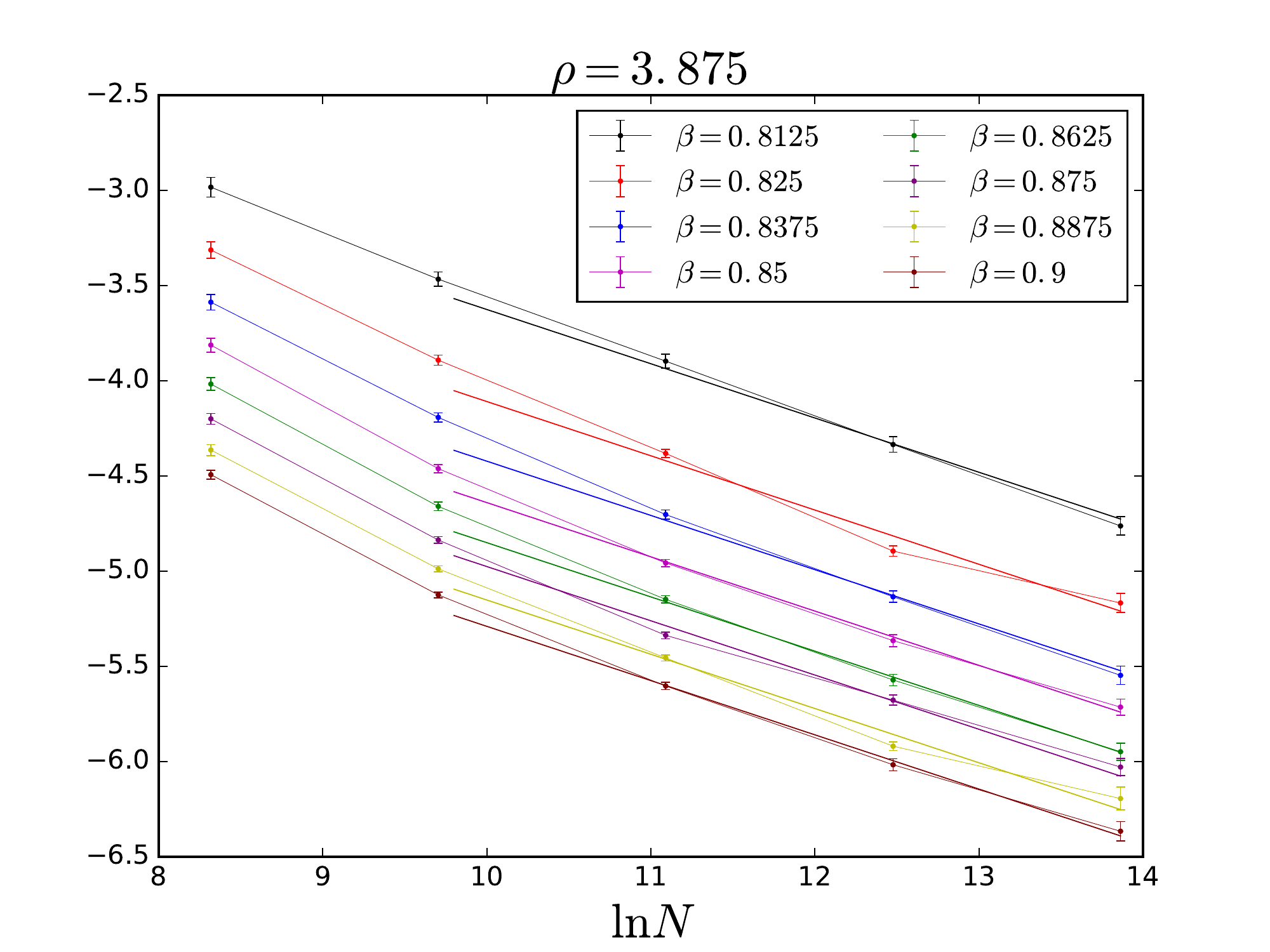}
	\caption{Left: $\ln (1-B(N,\beta))$ vs $\ln N$ for $\rho=4.5$. Different curves correspond to different values of $\beta$ in ${\cal B}_{4.5}$ (from bottom to top $\beta$ decreases). Lines are a guide to the eye. Right: idem, for $\rho=3.875$. The straight lines are the result of a single linear regression of the three larger sizes of all the temperatures, subject to a common slope among all $\beta$'s and a $\beta$-dependent intercept. The total sum of squared residuals per degree of freedom is lower than one. \label{fig:bindervsL}}
\end{figure}

In figs. \ref{fig:bindervsbeta} and \ref{fig:bindervsL} we show our estimations of $B(N,\beta)$, versus $\beta$ for several values of $N$, and versus $N$ for several values of $\beta$, respectively, for both $\rho=4.5$ and $\rho=3.875$. Our reference simulation at $\rho=4.5$ exhibits an overlap, under the statistical errors, of the curves  $B(N,\beta)$ for large $N$'s at under-critical temperatures, as expected (see section \ref{sec:methods}), since the superposition is a signature of the BKT universality class. Such a superposition is absent in the $\rho=3.875$ case, exhibting, instead, the typical crossing in continuous order transition. 

The presence/absence of superposition is shown more explicitly in fig. \ref{fig:bindervsL}, in which we report $\ln (1-B(N,\beta))$ as a function of $\ln N$ for different values of $\beta>\beta^*$. In the $\rho=4.5$ case, the values of $B(N,\beta)$ corresponding to the three larger sizes $N=256^2,512^2,1024^2$ coincide within their statistical errors for lower values of $\beta$ or, at most, they decrease with a functional dependence on $N$ which is {\it slower} than power-law, suggesting that $B(\infty,\beta)<1$. Oppositely, for $\rho=3.875$,  the behavior of $1-B(N,\beta)$ is consistent with a decay to zero as a power law for all $\beta>\beta^*$, strongly suggesting that $\lim_{N\to \infty} B(N,\beta)=1$, and that the $B(N,\beta)$ curves cross at a common value $\beta^*$ for sufficiently large $N$, as expected in second-order phase transitions.

The regression in fig. \ref{fig:bindervsL} is a single linear fit $\ln (1-B(N,\beta))=-x_B\ln N + g_\beta$ of all the datasets, such that the slope $-x_B$ is forced to be a common parameter for all the temperatures and the intercept $g_\beta$ is free to depend on the temperature. The regression provides a slow dependence of $1-B$ with the size: $x_B=-0.253(2)$, i.e., roughly the inverse of the square root of the linear size $L=N^{1/2}$. In appendix \ref{sec:binderNscaling} it is shown that such a slow dependence can be understood under the assumption of a dependence on $N$ of $\<|{\bf m}|\>$ of the type $\<|{\bf m}|\>(N,\beta) = \mu_{\infty}(\beta) + c_{\rho,\beta} N^{-x}$ with $x\simeq 1/8$, a dependence for which we also provide numerical evidence in appendix \ref{sec:binderNscaling}.  

We conclude that the data regarding the Binder cumulant indicates that the $\rho=3.875$ dilute XY model does not exhibit scale invariance in an interval of under-critical temperatures, contrary to the $\rho=4.5$ system.

\section{Conclusions\label{sec:conclusions}}
\label{sec:conclusions}

MC simulations of the $\rho=3.875$ long-range diluted XY model, based on the analysis of the Binder cumulant and on the modulus of the magnetization, clearly indicate that this model does present spontaneous magnetisation, and does not present scale invariance in an interval of under-critical temperatures. Consequently, it does not seem to belong to the BKT universality class, at least for what concern these two features. In its turn, according to the arguments of section \ref{sec:context}, this suggest that the limit power of the BKT universality class interval $[\rho_{\rm sr}:\infty]$ is $\rho_{\rm sr}=4$, in agreement with the results of \cite{defenu2015} or, more safely, that it is $\rho_{\rm sr}>3.875$. 

These results are also a numerical confirmation of the direct and inverse generalised Mermin-Wagner theorem for the XY model on graphs.\cite{cassi1992,burioni1999} Moreover, they suggest that the XY model with long-range interactions, according to the graph-long range equivalency discussed in section \ref{sec:intro}, exhibits at least some features of the BKT universality class as far as the power of the interaction decay is $\rho>4$, and vice-versa.

As side results, we have observed that: (1) the modulus of the magnetisation of systems with $\rho>4$ decreases as a power law $[\<|{\bf m}|\>](N,\beta)\sim N^{-a(\beta)}$ for under-critical temperatures (an estimation of the power $a(\beta)$ is reported in fig. \ref{fig:absm} for various values of $\beta$). (2) The critical inverse temperatures of the systems with $\rho=3.875$ and $\rho=4.5$, by inspection of the Binder cumulant crossing and overlapping, respectively, are, roughly, $\beta^*_{3.875}\sim 0.795(5)$ and  $\beta^*_{4.5}\sim 0.962(5)$, in agreement with those found in ref.\cite{berganza2013}. Furthermore, a rough estimation of the residual spontaneous magnetisation $\mu_\infty$ in the thermodynamic limit for $\rho=3.875$ as a function of $\beta$ is presented in fig. \ref{fig:absmfit3875}. We argue that $\<|{\bf m}|\>=\mu_\infty(\beta) + c(\beta) N^{-x}$ with $x\simeq 1/8$ in this case, and we relate the exponent $x$ to that governing the finite-size dependence of the Binder cumulant in the ferromagnetic phase, $1-B\sim N^{-x_B}$ with $x_B\simeq 1/4$.

As remarkable methodological side results, of which we provide detailed explanations in the appendices, we have observed that: (1) the observables depending explicitly on the magnetisation components $m_{x}$, $m_y$, and not on the modulus of the magnetisation $|{\bf m}|$, do not reach thermal equilibrium in simulation time scales, not even for moderate sizes $N$ and for the particularly low algorithmic cost of our parallel software. (2) For a correct estimation of the observable uncertainty, it is necessary to account for the fluctuations induced by the average over the ensemble of diluted graphs $[\cdot^2]-[\cdot]^2$. Such fluctuations are larger than the thermal fluctuations, specially for graphs near the short-range threshold  $\rho\sim \rho_{\rm sr}$, and for critical and under-critical temperatures (see fig. \ref{fig:errorcomparison}). In appendix \ref{sec:graphgeneration} we provide two numerical recipes to account for the graph-to-graph fluctuations and consequent statistical errors, and for their increment with respect to thermal fluctuations.

Possible completions of this work are the calculation of the critical exponents from our simulated data by means of the quotient method\cite{amit2005} and an investigation of the relationship of these results with the observation of the so called supra-oscillating dynamical phase of the XY in graphs.\cite{denigris2013,denigris2013a,denigris2013b,denigris2015,expert2017} It is possible that such a state, presenting particularly long correlation times, corresponds to graphs with  $d_{\rm s}\simeq 2$ (or $\rho\sim 4$), thus making evident the link between such a phenomenology and the results of the present work.

\begin{acknowledgments}
We are thankful to Alessandro Codello, Nicol\`o Defenu, Luca Leuzzi, Federico Ricci-Tersenghi and Stefano Ruffo for their useful comments. Special thanks to Andrea Maiorano and Andrea Trombettoni for their methodological and bibliographical suggestions. We  gratefully  acknowledge  the  support  of  NVIDIA Corporation  with  the  donation  of  the  Tesla  K40  GPU used for this research.
\end{acknowledgments}

\appendix
\section{Error estimation and correlation times \label{sec:errors}}

In this section we will refer to {\it realisation} a single implementation of the MC Markov chain with given initial conditions, sequence of random numbers and realisation of the graph.  The average over realisations is denoted by $[\cdot]$, while the average over the $n_{\rm meas}$ {\it temporal measures} within a single realisation is denoted by $\<\cdot\>$.

\subsection{Jack-Knife error estimation in a single realisation}

Within a realisation, the observables at different times are, in general, correlated, so that the statistical error associated to the average of an observable $\<\O\>$ is larger than the naive standard deviation of the sequence of measures, $\sigma_{\O}/{n_{\rm meas}}^{1/2}$. We now explain the way in which the uncorrelated error, $\sigma_r$, of the ensemble average within the $r$-th realisation, $\<\O\>_r$ (or, in general, in an arbitrary sequence of $n_{\rm meas}$ of observables), is estimated. 
We have used the Jack-Knife method for error estimation.\cite{amit2005,berg2004} The dataset $\{{\bf m}(t)\}_{t=1}^{n_{\rm meas}}$ of measures of the total magnetization is divided into {\it blocks} of size $b$. The Jack-Knife estimator for the variance of $\<\O\>$ (not of $\cal O$) is: 

\begin{equation}
	\sigma^2_{(b)}=\frac{1}{n_{\rm b}-1} \left[ \frac{1}{n_{\rm b}}\sum_{j=1}^{n_{\rm b}}\O_{(j,b)}^2-{\overline{\O_{(b)}}}^2 \right]
\end{equation}
where $j$ runs from 1 to the {\it number of blocks} $n_{\rm b}={\sf int}(n_{\rm meas}/b)$, $\O_{(j,b)}$ is the observable averaged over the set of magnetisations in the $j-$th block, and $\overline{\O_{(b)}}$ is the average over $j$ of the later quantity. 

For large blocksize $b$, the square root of the Jack-Knife variance, $\sigma_{(b)}$, does not depend on $b$ within its statistical error (see below)  and it becomes a fair error for $\<\O\>$. This estimation coincides with the one obtained from the more immediate Bootstrap method.\cite{amit2005,berg2004} However, the Jack-Knife method also allows for the estimation of the correlation time associated to $\<\O\>$ (given $N$, $\beta$, and the algorithm used):\cite{amit2005,berg2004}

\begin{equation}
	\tau_{(b)} = \frac{{\sigma_{(b)}}^2}{\sigma^2_{\O}/n_{\rm MCS}}
	\label{eq:correlationtimes}
\end{equation}
where ${\sigma_{(b)}}^2$ is the Jack-Knife variance for large enough blocksizes $b$, and $\sigma^2_{\O}$ is the naive variance of the series of $\O(t)$ measurements from which the average $\<\O\>$ has been computed. The last equation has been used to estimate the correlation times that we present in the following subsection. 

We remark that the Jack-Knife method not only allows to estimate the standard deviation of correlated variables, but also to compute the error of an observable of the data $\O[\{{{\bf s}_i}\}_i]$, beyond linear propagation, and to estimate the error of the error (the error of $\sigma^2_{(b)}$, since it obeys a $\chi^2$-distribution\cite{berg2004}). We have used such an estimation for the error of the error in order to estimate, given $N$, $\beta$ and the observable, the minimum blocksize $b_{\rm min}$ such that $\sigma^2_{(b)}$ do not change within its errors for $b>b_{\rm min}$. Furthermore, the error of the error has been used to plot the error bars corresponding to the various correlation times, $\tau$, in the following section. 

Finally, the standard deviation associated to $\O$ within a single realisation $r$, $\sigma_r$, is estimated as $\sigma_{(b_{\rm min})}$.\footnote{Eventually, $\sigma_r$ may also be estimated as $\sigma_{(b_{\rm min})}+s(\sigma_{(b_{\rm min})})$, where $s(\sigma_{(b_{\rm min})})$ is the error of $\sigma_{(b_{\rm min})}$, obeying the $\chi^2$-distribution.}

\subsection{Intra-graph and inter-graph error estimation}

As we have anticipated, we have estimated the statistical uncertainty of the  average (over graph and time) of a generic observable, $[\<\O\>]$, in three different ways. 

\begin{enumerate}
	\item As the Jack-Knife error associated to different MC measures of a single graph realisation, $\sigma_r$, averaged over many realisations of the graph, $[\sigma_r]/N_{\rm r}^{1/2}$, where $N_{\rm r}$ is the number of realizations and where $[\cdot]=(1/N_{\rm r})\sum_r\cdot$. 
\item As the standard deviation of the single-realization measure $\<\O\>_r$ over different realizations, $([\<\O\>_r^2]-[\<\O\>_r]^2)^{1/2}/N_{\rm r}^{1/2}$.
\item As the Jack-Knife error associated to the MC measures in the {\it concatenated} series of MC simulations corresponding to different realisations, treating them as a single equilibrated simulation in which also the graph realisation has changed. 
\end{enumerate}

While (1) is the ensemble average, intra-graph error of $[\<\O\>]$ (averaged over graph realisations), (2,3) are expected to coincide and to account for both the inter-graph realisation and (intra-graph) ensemble error of $[\<\O\>]$. Indeed, in the particular case in which $[\cdot]$ is performed over many independent Markov Chain MC realisations with {\it the same graph}, both definitions are equivalent, and they are expected to coincide within their errors (of the error). For this reason, when {\it different} graph realisations are averaged in $[\cdot]$, the (2,3) errors are larger or equal than (1) and the excess of (2,3) with respect to (1) accounts for the amount of  the error due from the inter-graph realisation only.

\begin{figure}
\includegraphics[width=.475\columnwidth]{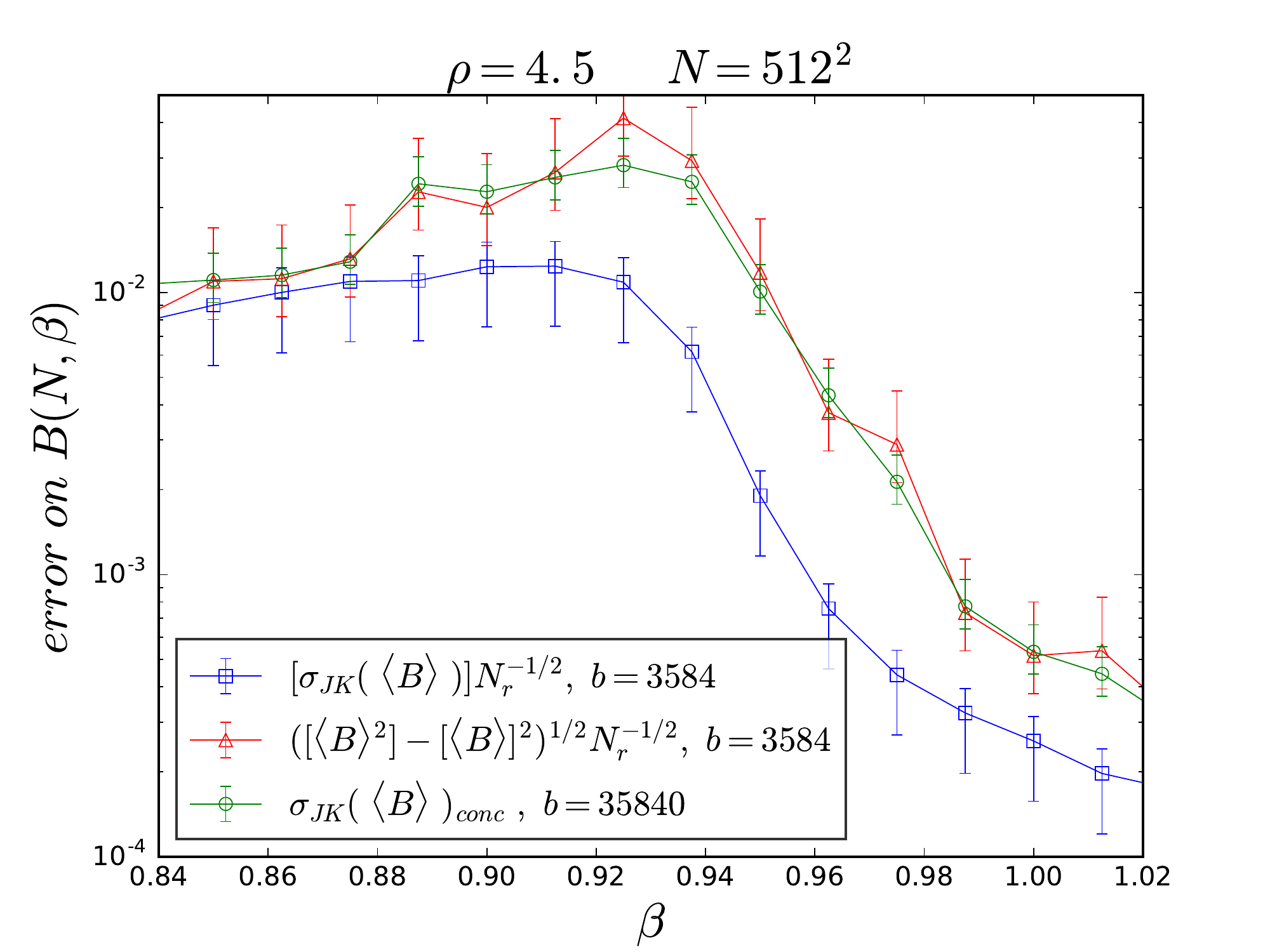}
\includegraphics[width=.475\columnwidth]{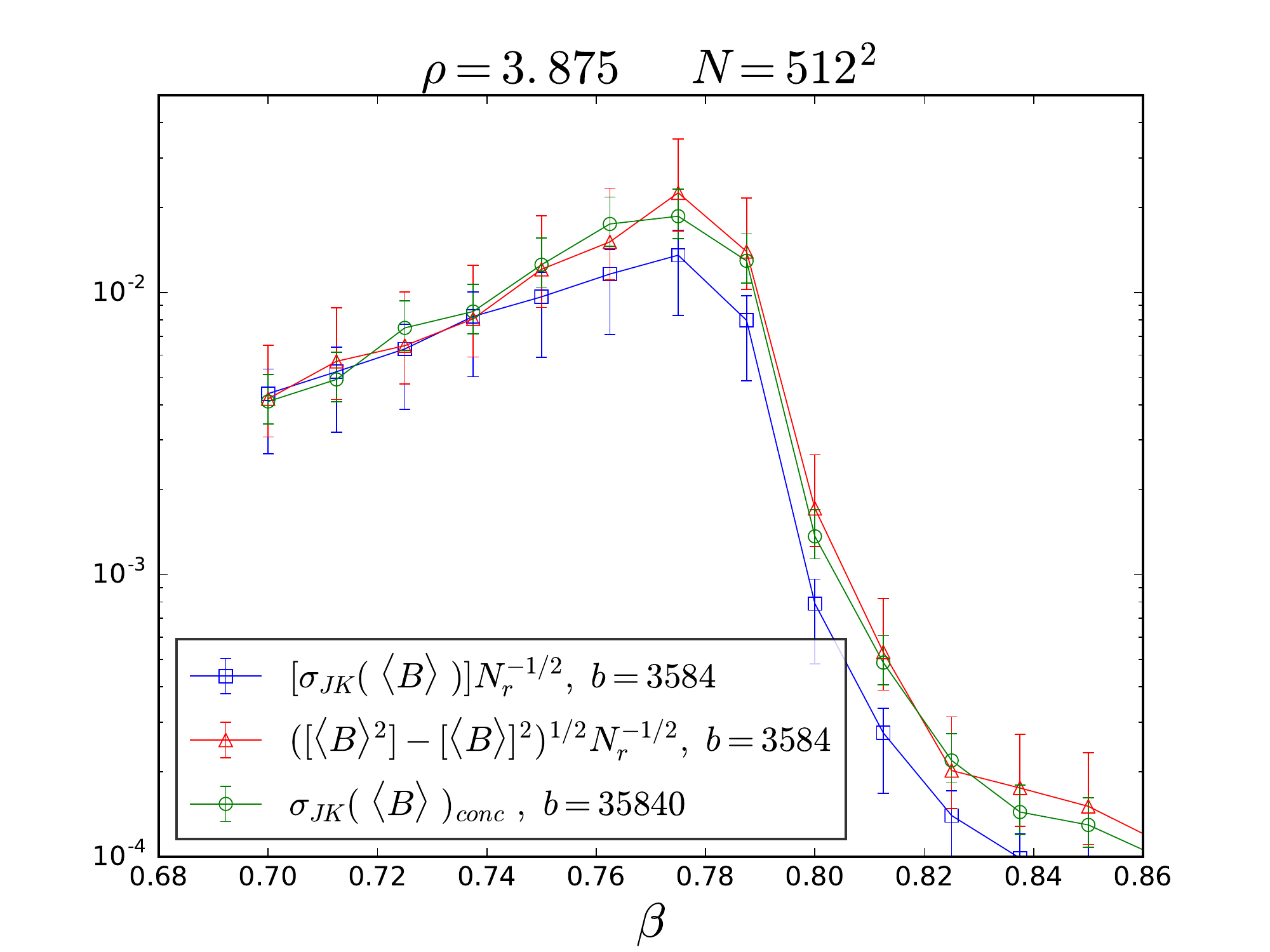}
	\caption{Left: Intra-graph and inter-graph error of the Binder cumulant versus $\beta$ for $\rho=4.5$, $N=512^2$, according to the three methods (1-3) discussed in the text. The error bar corresponds to the confidence interval at the $97.5\%$ according to the $\chi^2$-distribution with number of  degrees of freedom $m=N_{r}$, $m=n_{\rm b}=n_{\rm MCS}/b$ and $m=n_{\rm b}=n_{\rm MCS} N_{\rm r}/b$, respectively. Right: idem, for $\rho=3.875$  \label{fig:errorcomparison}}
\end{figure}

In fig. \ref{fig:errorcomparison}, we show a comparison between the three errors for the Binder cumulant $\O=B$, $N=512^2$, $\rho=4.5$ and different values of $\beta$, in abscissa. At large temperatures, the fluctuations become independent of the graph realisation and the three errors coincide, while in the scaling region around the finite-size critical temperature and below the critical temperature, the influence of the graph is non-negligeable.  We conclude that the error (1), accounting only for the thermal fluctuations is an underestimation of the error in these kind of systems, in which the graph topological disorder induces nontrivial inter-graph fluctuations. As error bars in all the figures of this article we have consequently considered the error (3).

\subsection{Correlation times}

\begin{figure}
\includegraphics[width=.65\columnwidth]{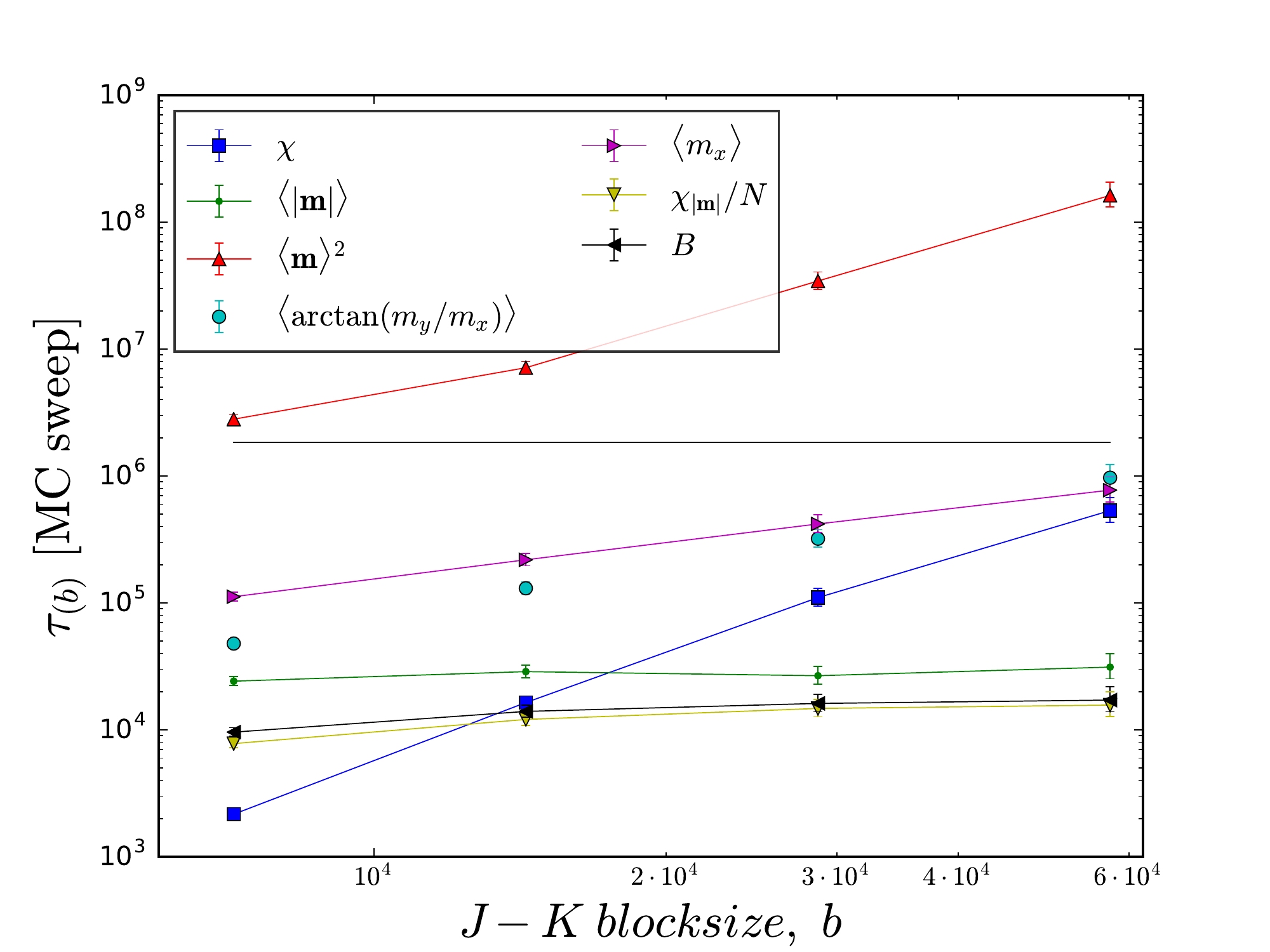}
	\caption{Estimated correlation times $\tau_{(b)}$ of various averaged observables $\cal O$ in MCS units as a function of the block-size of the Jack-Knife algorithm, $b$, for $\rho=4.5$, $N=512^2$, $\beta=1.05$. Lines are a guide to the eye. The horizontal line signals the number of MCS used in our simulations, $n_{\rm MCS}=1.835\cdot 10^{6}$. \label{fig:correlationtimes}}
\end{figure}

 We have estimated the correlation times of different observables by means of equation (\ref{eq:correlationtimes}) and using a dedicated, particularly long simulation with $\rho=4.5$, $N=512^2$, $\beta=1.05$, $n_{\rm MCS}\simeq 1.34\cdot 10^8$. In particular, we present $\tau_{(b)}$ vs $b$ in fig. \ref{fig:correlationtimes}, for a series of observables: the susceptibility, the modulus square of the magnetisation, the single component of the magnetisation, the argument of the magnetisation, the modulus of the magnetisation, the susceptibility of the modulus of the magnetisation, the Binder cumulant:

 \begin{eqnarray}
	 \chi/N &=& [\<{\bf m}^2\>-\<{\bf m}\>^2] \\
	 {\bm \mu}^2&=&[\<{\bf m}\>^2] \\
	 \mu_x&=&[\<{m_x}\>]\\
	 \phi&=&[\<\arctan(m_y/m_x)\>] \\
	 \mu&=&[\<|{\bf m}|\>]\\
	 \chi_{|{\bm m}|}/N&=& [\<{\bf m}^2\>-\<|{\bf m}|\>^2]  \\
	 B&=&2-\left[\frac{\<{\bf m}^4\>}{\<{\bf m}^2\>^2}\right]
 \end{eqnarray}

As it is apparent from fig. \ref{fig:correlationtimes}, the correlation times of the observables explicitly depending on the single components $m_{x,y}$, through a functional dependence different from the modulus, $|\bf m|$, exhibit very large values of $\tau_{(b)}$, much larger than the simulation time $n_{\rm MCS}\sim 2\cdot 10^{6}$ and, most importantly, they increase monotonously  with the block size in a statistically significant way, indicating that $\tau_{(b)}$ (and, consequently, the error $\sigma_{(b)}$) is an underestimation of the real correlation time and, consequently, {\it that the observable is not thermalised}. Oppositely, the observables depending on $|\bf m|$ only, exhibit a value of $\tau_{(b)}$ which is constant in $b$ within their errors, and that is lower than the simulation time. 

The correlation time of the observable $|{\bf m}|$ is moderate ($\tau\ll 10^{5}$): independently of the average global {\it angle} of the magnetisation vector, the system quickly acquires the equilibrium value of $\<|{\bf m}|\>$. However, the Hamiltonian is invariant under global rotations of all the spins, implying an enormous correlation time of the observable $\phi$ and of all observables explicitly depending on the single components of the magnetization, whose error, proportional to $\tau_{(b)}$ (\ref{eq:correlationtimes}), is underestimated.  This prevents equilibrium measurements of the observables $\chi$ and $\<{\bf m}\>$, at least in this circumstances and for the used algorithms. 

For the rest of the observables, presenting a low correlation time, the effective number of uncorrelated measures (given an observable, $N$, $\beta$ and $\rho$) is of the order of $n_{\rm meas}$ divided by the correlation time in units of $t_{\rm s}$, or $n^{\rm meas} t_{\rm s}/\tau_{(b)}$ (see eq. \ref{eq:correlationtimes}).

\section{Efficient generation of diluted long-range graphs\label{sec:graphgeneration}}

In order to perform accurate estimations of observables presenting large correlation times, as it is the case of the two-dimensional long-range diluted XY model with $\rho\simeq 4$, an accurate estimation of the error of the relevant quantities is crucial. We have already seen that, for an accurate estimation of the error, it is necessary to account for inter-graph fluctuations  (see appendix \ref{sec:errors}), hence to perform averages over different realizations of the graph.

With this aim, we have implemented an improved version of the algorithm used (and described in \cite{berganza2013}) for the generation of single realizations of the long-range diluted graph. For large values of $\rho$, the original algorithm cost is $\ordine{N^2}$. Indeed, the algorithm starts from a graph with zero links, it considers a given random potential link; the link is finally added with its probability, $\sim |{\bf r}_{ij}|^{-\rho}$. For large $\rho$, only nearest neighbors of the reference two-dimensional lattice present a non-negligeable probability of being created. Since the fraction of nearest neighbors is $\ordine{N^{-1}}$, and the algorithm adds $N_l=2N$ of them, the complexity becomes $\ordine{N^2}$.

The improved algorithm first calculates and stores a probability distribution of a link exhibiting distance $d=|{\bf r}_{ij}|$, $P(d) = d^{-\rho} g(d)/Z$, where $g(d)$ is the multiplicity of $d$ in the reference lattice with the given boundary conditions, and $Z$ is the corresponding normalizing factor. Afterwards, the following set of operations is sequentially performed, $N_l$ times: a distance, $d$, is extracted with its relative probability, $P(d)$; a lattice site $i$ is then chosen at random; a random neighbor, $j$, of $i$, is chosen at random among those satisfying the constraint $|{\bf r}_{ij}|=d$, considering the given boundary conditions; the link $ij$ is, then, created. The neighbor $j$ is taken from a lookup table, and the construction of the list with all the possible distances of the lattice is performed efficiently with the help of a hashing algorithm allowing for sub-linear search in the list (to check whether a given distance is already present in the list, in this way avoiding a further $\sim\ordine{N^2}$ cost). The algorithm has been checked comparing the indistinguishability of the degree histograms of the old and new algorithms for various values of $\rho$, $N$. 

\begin{figure}
\includegraphics[width=.65\columnwidth]{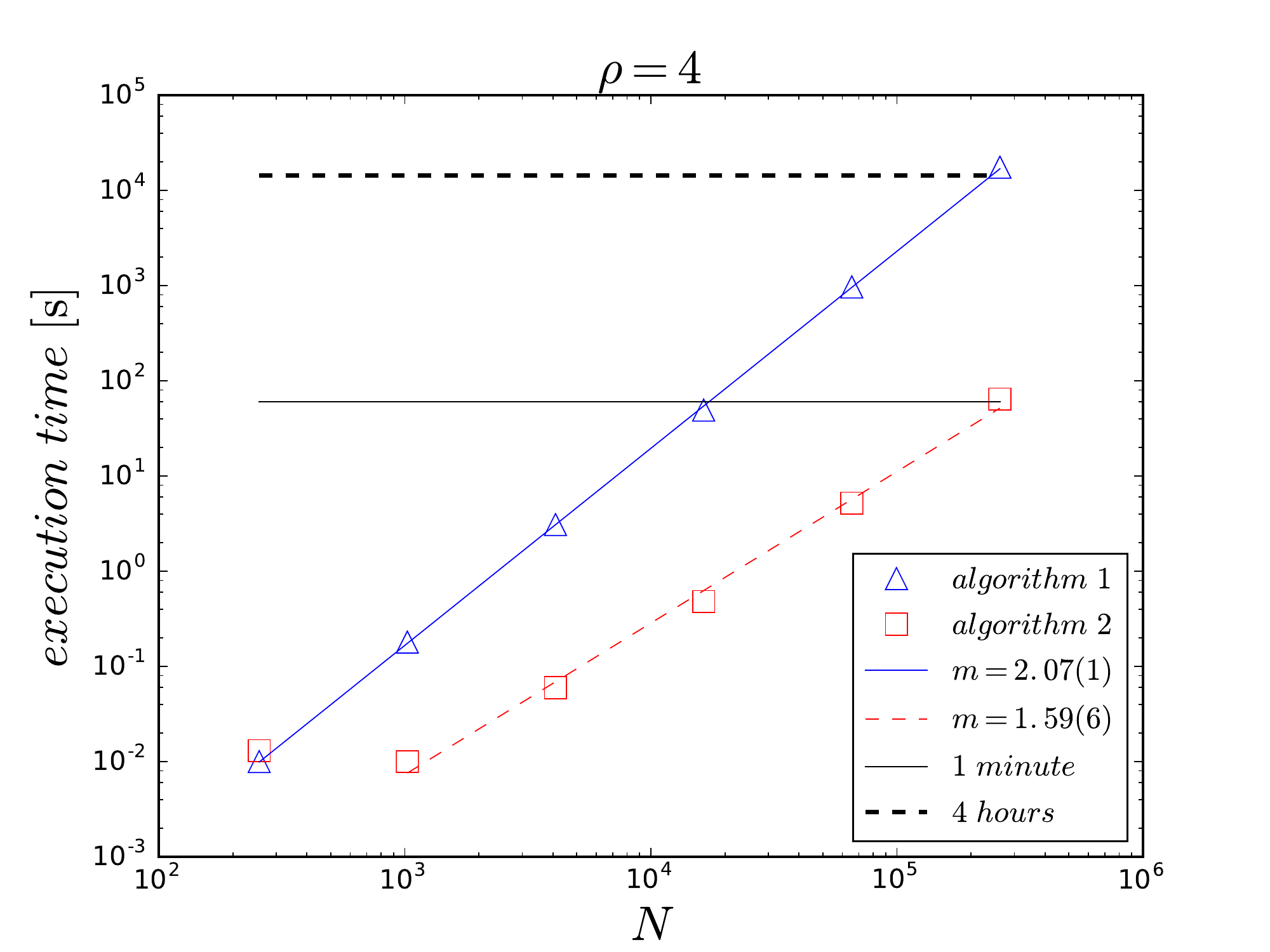}
	\caption{Execution time (seconds) required by generation of a $\rho=4$ diluted graph versus the size of the graph $N$, according to the 2012 algorithm (triangles) and to the novel version of the algorithm, ``algorithm 2'' (squares), using a 2.5GHz Intel-6 CPU. Straight lines are linear fits, whose slope is reported in the legend. \label{fig:graphgeneration}}
\end{figure}

The complexity of the new algorithm is over-linear, $\sim\ordine{N^{\Theta}}$, with $\Theta\simeq 1.6$, albeit both the generation of $P(d)$ and the choice among the $N$ neighbors at their chosen distance $d\leftarrow P(d)$ are performed in linear time. The over-linear time comes from the sorting of the list of possible distances according tho their probability $P(d)$, in its turn used to calculate the cumulative of $P(d)$ in order to extract $d\leftarrow P(d)$. Such an over-linear part, however, can be easily cured in successive improvements of the algorithm (by choosing a linear-time sorting algorithm). In any case, the new algorithm is much faster, allowing the generation of a $N=10^6$ graph in roughly one minute, against the $\sim$four hours of the former version (see fig. \ref{fig:graphgeneration}).

\section{Finite-size scaling of the Binder cumulant in the low-temperature phase \label{sec:binderNscaling}}

In sec. \ref{sec:scaleinvariance} we have shown that, in the $\rho=3.875$ case where the system is expected to exhibit a second-order phase transition, the Binder cumulant approaches its thermodynamic value, $1$, with a slow power-law dependence with the size: $1-B \sim N^{-x_B}$ with $x_B\simeq 1/4$. We will argue in this section that such a law is compatible with a dependence of the absolute value of the magnetisation with the size of the type:

\begin{equation}
	\label{eq:absmvsN}
	\<|{\bf m}|\>(N,\beta) = \mu_{\infty}(\beta) + c_{\rho,\beta} N^{-x}, \qquad x\simeq 1/8
\end{equation}
where $\mu_\infty(\beta)=\<|{\bf m}|\>(\infty,\beta)$, $c_{\rho,\beta}$ is a function of $\rho$ and $\beta$ only, and the exponent $x\simeq 1/8$ for $\rho=3.875$, independent of temperature. Afterwards, we will show that, indeed, the value $x\simeq 1/8$ is compatible with the results of our simulations.

We consider the XY model for $\rho<\rho^*$, when it presents a second-order phase transition. We assume that the probability distribution of the magnetisation in a single realisation of the graph is a product of two independent Gaussian distributions on its components, given a single realisation of the graph and a single initial condition: $p_N({\bf m}) = {\cal N}(m_x|\mu_x,\sigma^2) \times {\cal N}(m_y|\mu_y,\sigma^2)$. In the ferromagnetic phase, $\mu_{x,y}$ are nonzero in general, while in the paramagnetic phase they vanish. The variance of the distribution is proportional to the susceptibility, $\sigma^2=\chi/(2N)$. We are interested in the finite-size scaling of the quantity:

\begin{equation}
	\frac{\<{\bf m}^4\>}{\<{\bf m}^2\>^2}=\frac{ \mu^4 + 8\sigma^4 + 8 \mu^2\sigma^2}{ \mu^4 + 4\sigma^4 + 4 \mu^2\sigma^2}
	\label{eq:BinderXY}
\end{equation}
where $\mu=|{\bm \mu}|=(\mu_x^2+\mu_y^2)^2$, as it results from Gaussian integration. In the thermodynamic limit, the Binder cumulant, eq. \ref{eq:bindercumulant}, takes the values $1$ and $0$ in the ferromagnetic and paramagnetic phases respectively. 

We will suppose that, in the low-temperature phase, the $N$-dependence of $\<|{\bf m}|\>$ is $\<|{\bf m}|\> = \mu_\infty(\beta) + c_{\beta} N^{-x}$. We will omit the $\beta$-dependence of these constants, as we have done with that of $\chi$. This results in:

\begin{eqnarray}
	\frac{\<{\bf m}^4\>}{\<{\bf m}^2\>^2} = \frac{ (\mu_\infty + c N^{-x})^4 +8(\chi/2)^2N^{-2} + 8(\chi/2)^2N^{-1}(\mu_\infty + c N^{-x})^2 }{ (\mu_\infty + c N^{-x})^4 +4(\chi/2)^2N^{-2} +4(\chi/2)^2N^{-1}(\mu_\infty + c N^{-x})^2 } \\
\end{eqnarray}

In the ferrogmagnetic phase $\mu_\infty > 0$, and it is the only term of order $1$ in $N$. Dividing the numerator and the denominator by $\mu_\infty$, it is: 

\begin{eqnarray}
	\frac{\<{\bf m}^4\>}{\<{\bf m}^2\>^2} = \frac{1 + A_1 }{1+ A_2 }  \sim 1 - A_1 + A_1^2 + A_2 + {\sf O}[A_1^3,A_2^2]  
\end{eqnarray}
where $A_{1,2}$ denote the numerator and denominator terms decreasing with $N$. The leading term in $N$ of this expression is in $-A_1 + A_1^2 + A_2$, it is positive and of order $\sim N^{-2x}$ (more precisely, it is: $\mu_\infty^6 c^2 N^{-2x}$). We obtain, hence, that the introduction of an $N$-dependence on $\mu$ leads to the following leading finite-size dependence of the Binder cumulant:

\begin{eqnarray}
	1-B \sim N^{-x_{B}} , \qquad  x_B=2x \label{eq:binderNscaling}
\end{eqnarray}
and where $x$ is such that $\<|{\bf m}|\> = \mu_\infty + c N^{-x}$. 

\begin{figure}
\includegraphics[width=.7\columnwidth]{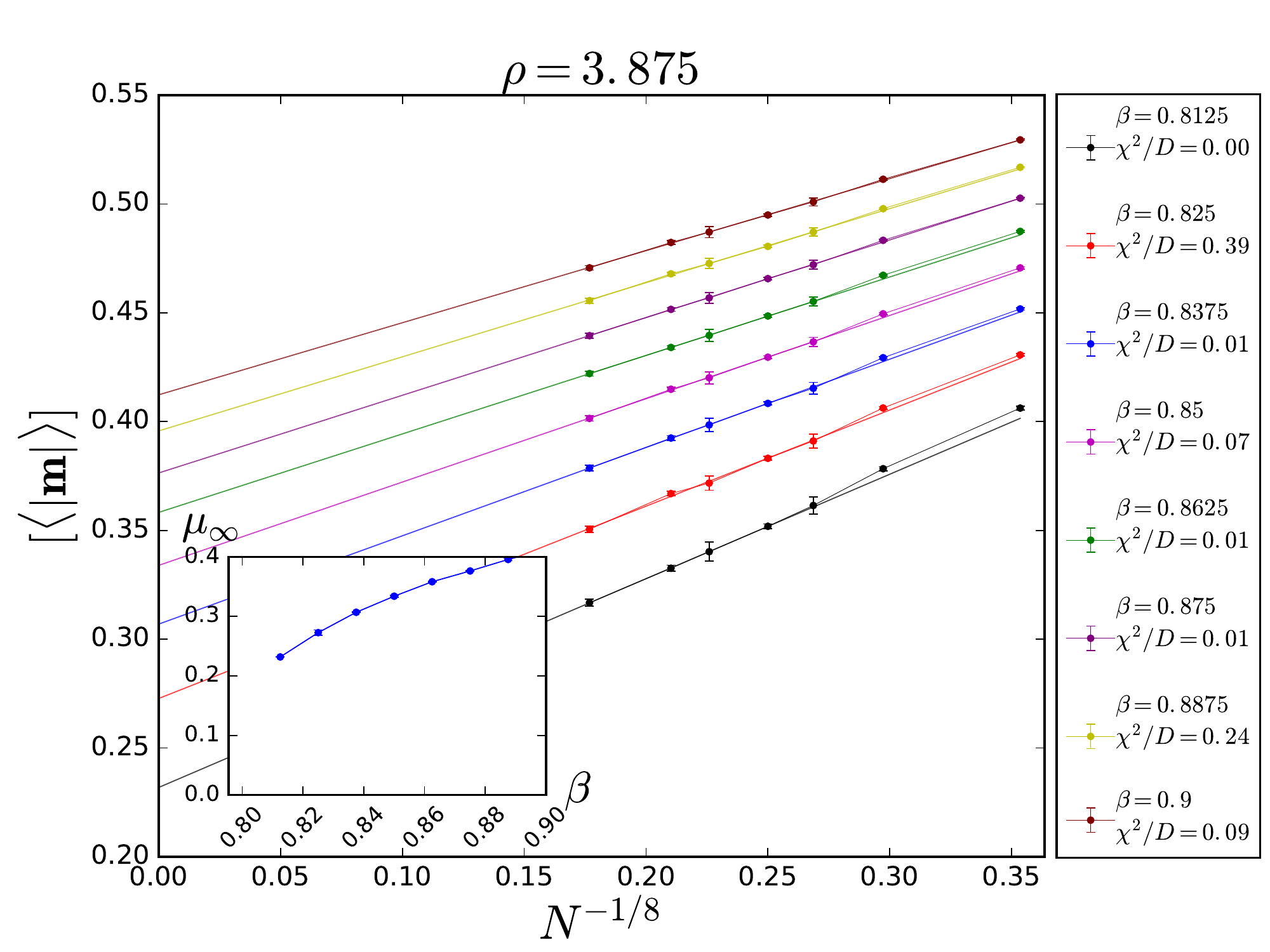}
	\caption{Main figure: $\<|{\bf m}|\>(N,\beta)$ versus $N^{-1/8}$ for all inverse temperatures in the interval ${\cal B}_{3.875}$ (from bottom to top $\beta$ increases). The straight lines are linear regressions whose squared sum of residuals per degree of freedom is reported in the legend. Inset: intercept of the fit, $\mu_\infty(\beta)$ for all inverse temperatures, vs $\beta$. The lines are a guide to the eye. The error is smaller than the point marker. \label{fig:absmfit3875}}
\end{figure}

The observed value of $x_B \simeq 0.25$ requires, hence, $x\simeq 1/8$. Let us now present some evidence supporting that, in the $\rho=3.875$ case, the average modulus of the magnetisation is, indeed, compatible with the law (\ref{eq:absmvsN}). This is shown in fig. \ref{fig:absmfit3875}, in which we show the average modulus of the magnetisation with respect to $N^{-1/8}$, along with a linear regression. The summed squared residuals per degree of freedom is lower than one for all the temperatures (in the $\rho=4.5$ case, this quantity is larger than $5$ for all the temperatures). The intercept of the fit allows to estimate the residual magnetisation $\mu_\infty(\beta)$, shown in the  inset of fig. \ref{fig:absmfit3875}, according to the assumption that $x=1/8$. 

We remark that $x\simeq 1/8$ is not a numerical estimation from the data but a guess with which the data is compatible, as shown by the low values of the squared residuals per degree of freedom in fig. \ref{fig:absmfit3875}. The error bars of $\<|{\bf  m}|\>$ are, however, large enough so that the data results compatible with other values of $x \lesssim 1/8$ as well. The estimation $x_B\simeq 1/4$ and the guess $x\simeq 1/8$ are, however, compatible with each other through the relation (\ref{eq:binderNscaling}), and both compatible with the data.

\bibliography{XYbib}

\end{document}